\documentclass[12pt,draftclsnofoot,onecolumn]{IEEEtran}

\usepackage{amsmath,amssymb,amstext,amsthm}
\usepackage{graphicx,color,subfig}
\usepackage{enumerate}
\usepackage{cite}
\usepackage{pgfplots,tikz}
\usepackage[americaninductors,americanvoltages]{circuitikz}
\newsavebox\LureNonlin

\newtheorem{thm}{Theorem}
\newtheorem{cor}{Corollary}

\newtheorem{prop}{Proposition}
\newtheorem{lem}{Lemma}
\newtheorem{rem}{Remark}

\newcommand{\bR}{\ensuremath{{\mathbb R}}}

\newcommand{\tl}{\ensuremath{\tilde}}
\newcommand{\al}{\ensuremath{\alpha}}
\newcommand{\be}{\ensuremath{\beta}}

\newcommand{\ga}{\ensuremath{\gamma}}
\newcommand{\del}{\ensuremath{\delta}}

\newcommand{\vDel}{\ensuremath{\varDelta}}
\newcommand{\ep}{\ensuremath{\epsilon}}
\newcommand{\ze}{\ensuremath{\zeta}}
\newcommand{\ka}{\ensuremath{\kappa}}
\newcommand{\Lam}{\ensuremath{\Lambda}}
\newcommand{\lam}{\ensuremath{\lambda}}
\newcommand{\vpi}{\ensuremath{\varpi}}
\newcommand{\vrho}{\ensuremath{\varrho}}
\newcommand{\si}{\ensuremath{\sigma}}
\newcommand{\Si}{\ensuremath{\Sigma}}
\newcommand{\vphi}{\ensuremath{\varphi}}

\newcommand{\cE}{\ensuremath{\mathcal{E}}}
\newcommand{\cG}{\ensuremath{\mathcal{G}}}

\newcommand{\cK}{\ensuremath{{\mathcal K}}}
\newcommand{\cL}{\ensuremath{{\mathcal L}}}

\newcommand{\cN}{\ensuremath{\mathcal{N}}}

\newcommand{\cX}{\ensuremath{\mathcal{X}}}
\newcommand{\ul}{\ensuremath{\underline}}
\newcommand{\ol}{\ensuremath{\overline}}

\newcommand{\sfK}{\mathsf{K}}

\newcommand{\diag}{{\rm diag}}

\allowdisplaybreaks

\begin{document}

\title{Adaptation and Disturbance Rejection for Output Synchronization of Incrementally Output-feedback Passive Systems}

\author{Hongkeun~Kim
        and~Claudio~De~Persis%,~\IEEEmembership{Member,~IEEE}%
\thanks{H. Kim is with School of Mechatronics Engineering, Korea University of Technology and Education, Cheonan-si 31253,  Korea (e-mail: hkkim@koreatech.ac.kr).}
\thanks{C. De Persis is with ENTEG, Faculty of Mathematics and Natural Sciences, University of Groningen, 9747 AG Groningen, The Netherlands (e-mail: c.de.persis@rug.nl).}}%
%\todo{Acknowledgments? Korean Science Foundation? I also have many projects to acknowledge... should we add them later?}}%

\maketitle

\begin{abstract}
	This note addresses the output synchronization problem of incrementally output-feedback passive nonlinear systems in the presence of exogenous disturbances.
	Two kinds of distributed controllers are proposed; one placed at the nodes and the other placed at the edges.
	Each of them is synthesized based on the adaptive control method to cope with the shortage of passivity, and on the internal model principle to deal with the disturbances.
	The proposed controllers synchronize the outputs of the nonlinear systems when the solution of the closed-loop system is bounded.
	Based on this, we present a class of systems for which boundedness of the solutions is guaranteed.
	The analysis used in this note is also applicable to a case where systems are coupled via links modeled by dynamical systems.
	Simulation results of a network of Van der Pol oscillators show the effectiveness of the proposed methods.
\end{abstract}

\begin{IEEEkeywords}
	Output synchronization, incremental output-feedback passivity, strong coupling, disturbance rejection, solution boundedness.
\end{IEEEkeywords}

%%%%%%%%%%%%%%%%%%%%%%%%%%%%%%%%%%%%%%%%%%%%%%%%%%%%%%%%%%%%%%%%%%%%%%%%%%%%%%%%%%%%%%%%%%%%%%%%%%%%%%%%%%%%%%%%%%%%%%%%%%%%%%%%%%%%%%%%%%%%%%
%%%%%%%%%%%%%%%%%%%%%%%%%%%%%%%%%%%%%%%%%%%%%%%%%%%%%%%%%%%%%%%%%%%%%%%%%%%%%%%%%%%%%%%%%%%%%%%%%%%%%%%%%%%%%%%%%%%%%%%%%%%%%%%%%%%%%%%%%%%%%%
\section{Introduction} \label{sec:Intro}
%%%%%%%%%%%%%%%%%%%%%%%%%%%%%%%%%%%%%%%%%%%%%%%%%%%%%%%%%%%%%%%%%%%%%%%%%%%%%%%%%%%%%%%%%%%%%%%%%%%%%%%%%%%%%%%%%%%%%%%%%%%%%%%%%%%%%%%%%%%%%%
%%%%%%%%%%%%%%%%%%%%%%%%%%%%%%%%%%%%%%%%%%%%%%%%%%%%%%%%%%%%%%%%%%%%%%%%%%%%%%%%%%%%%%%%%%%%%%%%%%%%%%%%%%%%%%%%%%%%%%%%%%%%%%%%%%%%%%%%%%%%%%

%\todo{I suggest to remove reference [3]; I am not convinced that the results there are correct.}
Over the past decades, synchronization of multiple systems interconnected through a graph has received considerable attention from the control community due to its large number of applications such as rendezvous, formation control, frequency synchronization of power grids, oscillators synchronization, to name a few.
Motivated by the passive properties of physical systems such as electrical networks and oscillators, the passivity-based approach has proven its usefulness in dealing with the synchronization problem.
See, e.g., \cite{Arcak07,Stan07a,Pogromsky01,Scardovi10,Shafi14,Bai13,Claudio14,Burger15} (refer to \cite{Burger14} for other approaches based on QUAD or convergent properties). 
In particular, important classes of nonlinear oscillators fall into the category of systems with a shortage of passivity for which proving synchronization still relies on passivity-based arguments \cite{Stan07a,Torres15}. 
 
When dealing with systems that exhibit a shortage of passivity, one of key requirements to achieve synchronization is the so-called strong coupling condition, meaning that the algebraic connectivity of the graph should exceed certain threshold value \cite{Stan07a,Pogromsky01,Scardovi10,Claudio14,Burger14} (even though there is no disturbance acting on the systems).
Another important requirement is that the solutions of the closed-loop system are bounded \cite{Shafi14,Bai13,Burger14}, which may not be guaranteed in general.
This is because Lyapunov functions used to show synchronization are often constructed from the differences between states and even when boundedness of the Lyapunov function is proven, it is not possible to infer boundedness of the states but only of their differences. 

Many dynamical networks are open systems that interact with the environment (e.g., an inventory system with in-/out-flow of material, a smart grid subject to unknown demand and generation). 
As such, they are often affected by external perturbations that can disrupt synchronization. Internal model based controllers have been recently proposed as a mean to restore synchronization despite the effect of disturbances \cite{Wei13,Claudio14,Burger14}. Previously, passivity-based internal model controllers were proposed to deal with formation control problem with unknown reference velocity \cite{Bai11}. Furthermore, passivity and internal model controllers play a role in the control of dynamical networks when constraints and optimality considerations must be also taken into account \cite{Burger14,Burger14a,Burger14b}. 

Motivated by the work discussed above, we study the output synchronization problem of incrementally output-feedback passive (iOFP) nonlinear systems on an undirected graph, in the presence of disturbances generated by exogenous systems.
Two different structures of distributed controllers are handled in this note.
In  the first one, each local controller is placed at the corresponding node, while in the second one, a controller is associated  to each edge.
Either at the nodes or at the edges, each controller is a combination of an adaptive law to cope with the shortage of passivity \cite[Chapter 6]{Khalil02} and internal model \cite{Byrnes97} to deal with the disturbances.
It is shown that despite the shortage of passivity and the external disturbances, the proposed controllers enforce the output synchronization of the iOFP systems, provided that the solution of the closed-loop system is bounded.
Then, we show that for nonlinear systems which are input-to-state stable (ISS) relative to a compact set $\cX$, boundedness of the solution can be guaranteed. 
A class of open-loop systems ensuring both iOFP and ISS properties is given as well.

The structure in which controllers are located at the nodes is common in the literature and it is already known that (iOFP) nonlinear systems interconnected via a {\em static} diffusive coupling achieve synchronization under strong coupling conditions \cite{Stan07a}. 
In this note, the difficulty descends from the fact that as opposed to the static diffusive coupling, the systems at the nodes are coupled via dynamical systems that aim at guaranteeing synchronization while rejecting the action of the disturbances.
Meanwhile, controllers at the edges have attracted interest more recently \cite{Burger14,Wei13, Burger15}. 
In certain problems in, e.g., distribution networks, the inputs to the dynamical systems at the nodes are constrained to satisfy certain physical laws (such as Kirchhoff's laws) and having controllers at the edges is more convenient since they regulate the ``flow" exchanged among the different nodes. 
In addition, the analysis carried out in the case of dynamical controllers at the edges turns out to be useful to deal with the case in which the dynamics at the edges is given, a feature which arises in those synchronization problems where the graph models physical interconnections among the nodes \cite{Torres15,Burger14a,Xiang14}.

A few remarks on the proposed controllers are as follows.
In contrast to \cite{Shafi14,Bai13}, each adaptive law assigned to the corresponding node (or edge) is one-dimensional, leading to lesser dimension of controllers in general.
Moreover, it does not require the symmetry (presented in, e.g., \cite{Shafi14,Bai13}) of the initial values and update gains of the adaptive laws.
Finally, the controllers are driven by relative outputs of the nonlinear systems only, while additional communication of the partial states of the internal models is necessary in \cite{Bai13}.
Thus, pure output feedback control for synchronization is achieved in our case.
%Finally, the scheme in this note is applicable to a case of systems coupled via dynamic links \cite{Torres15,Burger14a,Xiang14}.

%The rest of this technical note is organized as follows.
%Section \ref{sec:ProblemFormulation} presents the output synchronization problem that we consider.
%Two kinds of distributed controllers are proposed in Section \ref{sec:ContrDesign} with further discussions on the solution boundedness of the closed-loop system in Section \ref{sec:SolBoundedness}.
%As an example, a network of Van der Pol oscillators is considered in Section \ref{sec:Simulation} with some simulation results provided.
%A conclusion is drawn in Section \ref{sec:Conclusion}.

% NOTATION BEGIN -----------------------------------------------------------------------------------------------------------------------------

{\it Notation:}
We denote an undirected graph by $\cG = (\cN, \cE, A)$, where the node set $\cN = \{1, 2, \dots, N\}$ is a finite nonempty set of nodes, the edge set $\cE \subseteq \cN \times \cN$ is a set of pairs of distinct nodes satisfying $(i,j) \in \cE \Leftrightarrow (j,i) \in \cE$, and the adjacency matrix $A = [a_{ij}] \in \bR^{N \times N}$ is a symmetric nonnegative matrix defined in a way that $a_{ij} > 0$ if $(i,j) \in \cE$ and $a_{ij} = 0$ otherwise. 
A path connecting nodes $i \neq j$ is a sequence of distinct nodes, $\{ p_{1}, p_{2}, \dots, p_{d} \}$, such that $p_{1} = i$, $p_{d} = j$, and $( p_{g}, p_{g+1} ) \in \cE$.
In this case, the length of the path is $d - 1$.
An undirected graph is connected if for every pair of distinct nodes, there is a path connecting them.
The Laplacian matrix of $\cG$ is defined by $L := \vDel - A$, where $\vDel$ is the diagonal matrix whose $i$-th diagonal is $\vDel_i := \sum_{j \in \cN} a_{ij} = \sum_{j \in \cN} a_{ji}$ (by the symmetry of $A$).
The incidence matrix $B = [b_{ig}]\in \bR^{N \times E}$ of $\cG$, with $E := |\cE|$, is that $b_{ig} := - \sqrt{a_{ij}}$ and $b_{jg} := \sqrt{a_{ij}}$ for which $(i, j)$ is the $g$-th edge of the graph ($g = 1, \dots, E$), and $b_{ig} := 0$ otherwise.
By its construction, $L = BB^\top$ and $B^\top 1_N = 0$, where $\top$ and $1_N \in \bR^N$ denote the matrix transpose and the vector of ones, respectively. 
See, e.g., \cite{Godsil04} for the details of graph theory.

The stacking of vectors $x_1, \dots, x_N$ is denoted by $[x_1; \cdots; x_N]$.
$\diag(R_1, \dots, R_N)$ is the block diagonal matrix with its $i$-th diagonal block $R_i$.
The Kronecker product is denoted by $\otimes$ and $B^+$ denotes the Moore-Penrose pseudoinverse \cite{Ben-Israel03} of a matrix $B$.
Euclidean norm is denoted by $\| \cdot \|$, while the 1-norm is denoted by $\|\cdot\|_1$.
For a compact set $\cX$, we define $\| x \|_\cX := \min_{y \in \cX} \| x - y \|$.
$I_N$ is the identity matrix of dimension $N$ and $\Pi := I_N - (1/N) 1_N 1_N^\top$.
$\bR_{\geq 0}$ denotes the set of nonnegative real numbers.

% NOTATION END -------------------------------------------------------------------------------------------------------------------------------

%%%%%%%%%%%%%%%%%%%%%%%%%%%%%%%%%%%%%%%%%%%%%%%%%%%%%%%%%%%%%%%%%%%%%%%%%%%%%%%%%%%%%%%%%%%%%%%%%%%%%%%%%%%%%%%%%%%%%%%%%%%%%%%%%%%%%%%%%%%%%%
%%%%%%%%%%%%%%%%%%%%%%%%%%%%%%%%%%%%%%%%%%%%%%%%%%%%%%%%%%%%%%%%%%%%%%%%%%%%%%%%%%%%%%%%%%%%%%%%%%%%%%%%%%%%%%%%%%%%%%%%%%%%%%%%%%%%%%%%%%%%%%
\section{Problem Formulation} \label{sec:ProblemFormulation}
%%%%%%%%%%%%%%%%%%%%%%%%%%%%%%%%%%%%%%%%%%%%%%%%%%%%%%%%%%%%%%%%%%%%%%%%%%%%%%%%%%%%%%%%%%%%%%%%%%%%%%%%%%%%%%%%%%%%%%%%%%%%%%%%%%%%%%%%%%%%%%
%%%%%%%%%%%%%%%%%%%%%%%%%%%%%%%%%%%%%%%%%%%%%%%%%%%%%%%%%%%%%%%%%%%%%%%%%%%%%%%%%%%%%%%%%%%%%%%%%%%%%%%%%%%%%%%%%%%%%%%%%%%%%%%%%%%%%%%%%%%%%%

Let us consider a group of $N$ nonlinear systems, each of which is described by
\begin{align}
	\begin{split}
		\dot x_i &= f(x_i, u_i, d_i), \\
		y_i &= h(x_i), 
	\end{split} \qquad i = 1, 2, \dots, N, \label{eq:System}
\end{align}
where $x_i \in \bR^n$ and $u_i, y_i \in \bR^q$ are the state, input, and output of the $i$-th system, respectively, and is incrementally output-feedback passive (iOFP) in the sense that there exist a storage function $\Phi : \bR^n \times \bR^n \rightarrow \bR_{\geq 0}$, a number $\si \in \bR$, and two functions $\ul\al$ and $\ol\al$ of class $\cK_\infty$ such that $\ul\al(\| x_i - x_i' \|) \leq \Phi(x_i, x_i') \leq \ol\al(\| x_i - x_i' \|)$ and 
\begin{align}
	\frac{\partial \Phi}{\partial x_i} (x_i, x_i') f(x_i,u_i,d_i) &+ \frac{\partial \Phi}{\partial x_i'} (x_i, x_i') 
			f(x_i',u_i',d_i') \label{eq:Dissipation}\\
	&\leq \si (y_i - y_i')^\top (y_i - y_i') + (y_i - y_i')^\top \left( (u_i + d_i) - (u_i' + d_i') \right) \nonumber
\end{align}
hold for all $(x_i, u_i, d_i) \in \bR^n \times \bR^q \times \bR^q$ and $(x_i', u_i', d_i') \in \bR^n \times \bR^q \times \bR^q$.
Moreover, $f(\cdot,\cdot,\cdot)$ and $h(\cdot)$ are assumed to be locally Lipschitz and continuously differentiable in their arguments, respectively.
The system satisfying \eqref{eq:Dissipation} is often referred to as $\si$-relaxed cocoercive system \cite{Scardovi10}.
On the other hand, the system \eqref{eq:System} is said to be incrementally passive if $\si \leq 0$, and incrementally output-strictly passive if $\si < 0$.
Examples of iOFP systems include output-feedback passive linear systems, Goodwin oscillators (see, e.g., \cite[Example 1]{Claudio14}), and Van der Pol oscillators (see, e.g., Section \ref{sec:Simulation}).

The signal $d_i$ represents the disturbance acting on the $i$-th system and is generated by an exogenous system
\begin{align}
		\dot w_i &= s_i(w_i), \qquad d_i = R_i w_i,  \qquad w_i \in \bR^{m_i}, \label{eq:Exosystem}
\end{align}
satisfying that $s_i(0) = 0$ and for all $(w_i, w_i') \in \bR^{m_i} \times \bR^{m_i}$,
\begin{align}
	(w_i - w_i')^\top \left( s_i(w_i) -  s_i(w_i') \right) \leq 0. \label{eq:Exosystem_Dissipatioin}
\end{align}
Note that in this case, the solution of \eqref{eq:Exosystem} always exists and any ball centered at the origin is forward invariant for \eqref{eq:Exosystem}.
An example of such systems having the property \eqref{eq:Exosystem_Dissipatioin} is the one given by $s_i(w_i) = S_i w_i$ with $S_i$ skew-symmetric.

We assume that the $N$ systems \eqref{eq:System} are defined over a connected undirected graph $\cG = (\cN, \cE, A)$ with the node set $\cN = \{1, 2, \dots, N\}$.
Each node $i \in \cN$ is associated with the $i$-th system in \eqref{eq:System} and the edges in $\cE$ define the interconnection structure.
Under this setting, the objective is to design distributed controllers that enforce the asymptotic output synchronization of \eqref{eq:System}, i.e.,
\begin{align*}
	\lim_{t \to \infty} \| y_i(t) - y_j(t) \| = 0, \qquad i,j \in \cN.
\end{align*}
The difference of our contribution from the vast majority of the literature on synchronization is that synchronization should be guaranteed despite the presence of exogenous time-varying signals which are different from node to node and which introduce a source of heterogeneity in the network.

%%%%%%%%%%%%%%%%%%%%%%%%%%%%%%%%%%%%%%%%%%%%%%%%%%%%%%%%%%%%%%%%%%%%%%%%%%%%%%%%%%%%%%%%%%%%%%%%%%%%%%%%%%%%%%%%%%%%%%%%%%%%%%%%%%%%%%%%%%%%%%
%%%%%%%%%%%%%%%%%%%%%%%%%%%%%%%%%%%%%%%%%%%%%%%%%%%%%%%%%%%%%%%%%%%%%%%%%%%%%%%%%%%%%%%%%%%%%%%%%%%%%%%%%%%%%%%%%%%%%%%%%%%%%%%%%%%%%%%%%%%%%%
\section{Controller Design for Output Synchronization} \label{sec:ContrDesign}
%%%%%%%%%%%%%%%%%%%%%%%%%%%%%%%%%%%%%%%%%%%%%%%%%%%%%%%%%%%%%%%%%%%%%%%%%%%%%%%%%%%%%%%%%%%%%%%%%%%%%%%%%%%%%%%%%%%%%%%%%%%%%%%%%%%%%%%%%%%%%%
%%%%%%%%%%%%%%%%%%%%%%%%%%%%%%%%%%%%%%%%%%%%%%%%%%%%%%%%%%%%%%%%%%%%%%%%%%%%%%%%%%%%%%%%%%%%%%%%%%%%%%%%%%%%%%%%%%%%%%%%%%%%%%%%%%%%%%%%%%%%%%

In this note, we deal with two kinds of distributed controllers; one for controllers placed at the nodes (refer to, e.g., \cite{Shafi14,Bai13,Claudio14} or Section \ref{sec:ContrAtNodes}) with their inputs being
\begin{align}
	\rho_i = \sum_{j \in \cN} a_{ij} (y_j - y_i), \qquad i = 1, \dots, N, \label{eq:CouplingAtNode}
\end{align}
and the other for controllers placed at the edges (see, e.g., \cite{Burger14,Burger15} or Section \ref{sec:ContrAtEdges}) with their inputs constrained to be 
\begin{align}
	\vrho_g = \sum_{j \in \cN} b_{jg} y_j, \qquad g = 1, \dots, E. \label{eq:CouplingAtEdge}
\end{align}
In both cases, the algebraic connectivity (say $\lam_2$, the second smallest eigenvalue of $L$) of the graph $\cG$, and internal model principle \cite{Byrnes97} play crucial roles for the synchronizability of the systems \eqref{eq:System} as shown in \cite{Bai13,Burger14}.
Based on this observation, we provide internal-model-based solutions to the problem in the next subsections, where the strong coupling condition $\lam_2 > \si$ is no longer required.
This relaxation is done by using one-dimensional adaptive controller at each node (Section \ref{sec:ContrAtNodes}) or at each edge (Section \ref{sec:ContrAtEdges}).

%%%%%%%%%%%%%%%%%%%%%%%%%%%%%%%%%%%%%%%%%%%%%%%%%%%%%%%%%%%%%%%%%%%%%%%%%%%%%%%%%%%%%%%%%%%%%%%%%%%%%%%%%%%%%%%%%%%%%%%%%%%%%%%%%%%%%%%%%%%%%%
\subsection{Controllers at the nodes} \label{sec:ContrAtNodes}
%%%%%%%%%%%%%%%%%%%%%%%%%%%%%%%%%%%%%%%%%%%%%%%%%%%%%%%%%%%%%%%%%%%%%%%%%%%%%%%%%%%%%%%%%%%%%%%%%%%%%%%%%%%%%%%%%%%%%%%%%%%%%%%%%%%%%%%%%%%%%%

Homogeneous incrementally output-feedback systems of the form $\dot x_i =f(x_i) + u_i$ are known to synchronize for $u_i=  \sum_{j \in \cN} a_{ij} (x_j - x_i)$ under suitable conditions \cite{Stan07a}. 
This is not guaranteed any longer if disturbances $d_i$ are affecting the dynamics at the nodes. 
The controllers we propose in this section are a natural dynamical extension of the static diffusive coupling, in which the dynamics is introduced to deal with the heterogeneity of the disturbances.

Let us define $\bar x := \frac{1}{N} \sum_{j \in \cN} x_j$ and $\bar y := \frac{1}{N} \sum_{j \in \cN} y_j$.
Then, we have the following.

% THEOREM BEGIN: Controllers at Nodes --------------------------------------------------------------------------------------------------------

\begin{thm} \label{thm:ContrAtNodes}
	The outputs of the $N$ systems \eqref{eq:System} in closed-loop with the control
	\begin{subequations} \label{eq:ContrAtNode}
		\begin{align}
			\dot\xi_i &= s_i(\xi_i) - R_i^\top \rho_i, \label{eq:ContrAtNode_IntModel}\\
			\dot k_i &= \ga_i \rho_i^\top \rho_i, \qquad \qquad ~~\! \ga_i > 0, \label{eq:ContrAtNode_Adaptation} \\
			u_i &= - R_i \xi_i + k_i \rho_i, \qquad i \in \cN \label{eq:ContrAtNode_Output}
		\end{align}
	\end{subequations}
	synchronize asymptotically whenever the closed-loop solution is bounded.
	In particular, $\lim_{t \to \infty} \| y_i(t) - \bar y(t) \| = 0$ holds for $i \in \cN$.
\end{thm}

% THEOREM END: Controllers at Nodes ----------------------------------------------------------------------------------------------------------

% REMARK BEGIN: Lesser Dimension -------------------------------------------------------------------------------------------------------------

\begin{rem} \label{rem:LesserDimension}
	A different approach to cope with the strong coupling condition is to use adaptive laws corresponding to the edges as in \cite{Bai13,Shafi14}, e.g., by employing, instead of \eqref{eq:CouplingAtNode} and \eqref{eq:ContrAtNode_Adaptation}, $\rho_i = \sum_{j \in \cN} k_{ij} a_{ij} (y_j - y_i)$ and $\dot k_{ij} = \ga_{ij} (y_i - y_j)^\top (y_i - y_j)$ with
	\begin{align}
		k_{ij}(0) = k_{ij}(0), \qquad\quad \ga_{ij} = \ga_{ji} > 0. \label{eq:ContrAtNode_SymmetryCondition}
	\end{align}
	On the contrary, the update law \eqref{eq:ContrAtNode_Adaptation} is assigned to each node without the symmetry condition \eqref{eq:ContrAtNode_SymmetryCondition}.
	Thus, the number of adaptive laws in the entire system is less than that of \cite{Bai13,Shafi14} in general, yielding lesser dimension of controllers.
	In addition, only relative outputs of \eqref{eq:System}, aggregated into the variable $\rho_i$ in \eqref{eq:CouplingAtNode}, are required to be measurable to implement \eqref{eq:ContrAtNode}, while additional communication of partial states of th internal models \eqref{eq:ContrAtNode_IntModel} is necessary in the case of \cite{Bai13}.
\end{rem}

% REMARK END: Lesser Dimension ---------------------------------------------------------------------------------------------------------------

% PROOF BEGIN: Controllers at Nodes ----------------------------------------------------------------------------------------------------------

\begin{IEEEproof}
	Let us consider the function $V_1$ given by
	\begin{align}
		V_1(x) = \frac{1}{2} \sum_{i \in \cN} \sum_{j \in \cN} a_{ij} \Phi(x_i, x_j), \label{eq:V1}
	\end{align}
	where $x := [x_1; \cdots; x_N] \in \bR^{Nn}$.
	We note that by Lemma \ref{lem:V1_Positivity} in the Appendix, there are $\ul\eta$ and $\ol\eta$ of class $\cK_\infty$ such that $\ul\eta( \| \tl x \| ) \leq V_1(x) \leq \ol\eta( \| \tl x \| )$, in which $\tl x := (\Pi \otimes I_n) x = x - (1_N \otimes \bar x)$.
	Taking its time derivative along the solution of \eqref{eq:System} and using Lemma \ref{lem:CompactExpression} in the Appendix, we have
	\begin{align*}
		\dot V_1 (x) &\leq \frac{\si}{2} \sum_{i \in \cN} \sum_{j \in \cN} a_{ij} (y_i - y_j)^\top (y_i - y_j) + \frac{1}{2} \sum_{i \in \cN} \sum_{j \in \cN} a_{ij} (y_i - y_j)^\top (u_i - u_j + d_i - d_j) \\
		&= \si y^\top (L \otimes I_q) y + y^\top (L \otimes I_q) u + y^\top (L \otimes I_q) d,
	\end{align*}
	where $y := [y_1; \cdots; y_N]$, $u := [u_1; \cdots; u_N]$, and $d = [d_1; \cdots; d_N]$.
	Since $u = - (KL \otimes I_q)y - R \xi$, it further becomes
	\begin{align*}
		\dot V_1 (x) &\leq y^\top \big\{ (\si L - LKL) \otimes I_q \big\} y - y^\top (L \otimes I_q) (R\xi - d),
	\end{align*}
	where $R := \diag(R_1, \dots, R_N)$, $\xi := [\xi_1; \cdots; \xi_N]$, and $K := \diag(k_1, \dots, k_N)$.
	
	Next, let us define $\tl\xi_i := \xi_i - w_i$ and $\tl k_i := k_i - k_\star$ with $k_\star \in \bR$ to be determined, and consider the function 
	\begin{align*}
		V_2(\tl\xi, \tl k) = \frac{1}{2} \sum_{i \in \cN} \tl\xi_i^\top \tl\xi_i + \sum_{i \in \cN} \frac{\tl k_i ^2}{2 \ga_i},
	\end{align*}
	where $\tl\xi := [\tl\xi_1; \cdots; \tl\xi_N]$ and $\tl k := [\tl k_1; \cdots; \tl k_N]$.
	Then, from the property \eqref{eq:Exosystem_Dissipatioin}, its time derivative becomes
	\begin{align*}
		\dot V_2(\tl\xi, \tl k) &= \sum_{i \in \cN} \Big\{ \tl\xi_i^\top \big( s_i(\xi_i) - s_i(w_i) \big) 
				- \tl\xi_i^\top R_i^\top \rho_i + \tl k_i \rho_i^\top \rho_i \Big\} \\
		&\leq - \tl d^\top \rho + \rho^\top (\tl K \otimes I_q) \rho = \tl d^\top (L \otimes I_q) y + y^\top (L\tl KL \otimes I_q) y,
	\end{align*}
	where $\rho := [\rho_1; \cdots; \rho_N]$, $\tl K := \diag(\tl k_1, \dots, \tl k_N) = K - k_\star I_N$, and $\tl d := R \tl\xi$. 
	
	We finally consider the Lyapunov function $V(x, \tl\xi, \tl k) := V_1(x) + V_2(\tl\xi,\tl k)$, whose time derivative is given by
	\begin{align*}
		\dot V(x, \tl\xi, \tl k) \leq - y^\top \Big( (k_\star L^2 - \si L) \otimes I_q \Big) y.
	\end{align*}
	Let $U = [\frac{1}{\sqrt{N}} 1_N ~~ Q]$ be an orthogonal matrix such that $U^\top L U = \Lam$, where $\Lam = \diag(\lam_1, \dots, \lam_N)$, $\lam_1 = 0$, and $\lam_i > 0$ for $i \neq 1$.
	Let us assume, with no loss of generality, that $\lam_2$ is the smallest nonzero eigenvalue of the Laplacian, and choose $k_\star > 0$ sufficiently large to satisfy $\ep := (k_\star \lam_2 - \si) \lam_2 > 0$.
	Then, noting that $QQ^\top = \Pi$ and $\Pi = \Pi^2$, one can get
	\begin{align}
		\dot V(x, \tl\xi, \tl k) \leq - \ep y^\top ( QQ^\top \otimes I_q ) y = - \ep \tl y^\top \tl y, \label{eq:ContrAtNode_dotV}
	\end{align}
	where $\tl y_i := y_i - \bar y$ and $\tl y := [\tl y_1; \cdots; \tl y_N] = (\Pi \otimes I_q) y$.
	Integrating both sides of \eqref{eq:ContrAtNode_dotV}, we have $\tl y(t) \in \cL_2$, i.e., square integrable.
	Since the solution is bounded, $\dot x_i(t)$ and hence, $\dot{\tl y}_i(t)$ are bounded.
	The application of Barbal\u{a}t lemma \cite{Tao97,Khalil02} proves the result.
\end{IEEEproof}

% PROOF END: Controllers at Nodes ------------------------------------------------------------------------------------------------------------

In the proof of Theorem \ref{thm:ContrAtNodes}, the function $V_1(x)$ in \eqref{eq:V1} is used to construct a valid Lyapunov function for the closed-loop system.
If the system \eqref{eq:System} admits a quadratic storage function, i.e., $\Phi(x_i,x_i') = (x_i-x_i')^\top P (x_i-x_i')$ with $P = P^\top > 0$, then by using Lemma \ref{lem:CompactExpression} in the Appendix, we see that $V_1(x) = x^\top (L \otimes P) x = \tl x^\top (L \otimes P) \tl x$ which coincides with the one in \cite[Theorem 4]{Li13b}.
Therefore, in the absence of $d$ (because \cite{Li13b} does not consider disturbances), our Lyapunov function can be considered as a nonlinear version of the one in \cite[Theorem 4]{Li13b}. 

As one can see from the proof of Theorem \ref{thm:ContrAtNodes}, the design \eqref{eq:ContrAtNode} provides some flexibility.
When there is no external disturbance (i.e., \eqref{eq:System} is replaced by $\dot x_i = f(x_i, u_i)$, $y_i = h(x_i)$ and \eqref{eq:Dissipation} holds with $d_i \equiv 0$ and $d_i' \equiv 0$), the controller \eqref{eq:ContrAtNode_Adaptation} with its output $u_i = k_i \rho_i$, instead of \eqref{eq:ContrAtNode}, solves the problem without requiring the strong coupling condition $\lam_2 > \si$ imposed in \cite{Stan07a,Pogromsky01,Scardovi10}.
On the other hand, if $\lam_2 > \si$, then the use of the adaptive controller \eqref{eq:ContrAtNode_Adaptation} can be avoided.
In this case, the controller is simply given by \eqref{eq:ContrAtNode_IntModel} and $u_i = - R_i \xi_i + \rho_i$.

%%%%%%%%%%%%%%%%%%%%%%%%%%%%%%%%%%%%%%%%%%%%%%%%%%%%%%%%%%%%%%%%%%%%%%%%%%%%%%%%%%%%%%%%%%%%%%%%%%%%%%%%%%%%%%%%%%%%%%%%%%%%%%%%%%%%%%%%%%%%%%
\subsection{Controllers at the edges} \label{sec:ContrAtEdges}
%%%%%%%%%%%%%%%%%%%%%%%%%%%%%%%%%%%%%%%%%%%%%%%%%%%%%%%%%%%%%%%%%%%%%%%%%%%%%%%%%%%%%%%%%%%%%%%%%%%%%%%%%%%%%%%%%%%%%%%%%%%%%%%%%%%%%%%%%%%%%%

In this subsection, we propose distributed controllers for synchronization, each of which is placed at the corresponding edge \cite{Burger15,Burger14}. 
The input to the controller placed at the $g$-th edge is given by \eqref{eq:CouplingAtEdge} and the input to the $i$-th system \eqref{eq:System} is constrained to be
\begin{align}
	u_i = - \sum_{g = 1}^E b_{ig} v_g, \qquad i = 1, \dots, N, \label{eq:OutputCouplingAtEdges}
\end{align}
where $v_g \in \bR^q$ is the output of the $g$-th controller.
With this structural constraint, let us define $H := (B^+ \otimes I_q) R$ and let $H_g \in \bR^{q \times m}$ be the $g$-th block row of $H = [H_1; \cdots; H_E]$, where $R = \diag(R_1, \dots, R_N)$ and $m := \sum_{i = 1}^N m_i$. 
This specific value of $H$ in fact descends from the fulfillment of the regulator equations and the internal model property, shown in \cite{Burger15}, which are necessary to solve the output agreement problem for incrementally passive nonlinear systems under the constraint \eqref{eq:OutputCouplingAtEdges}.
In particular, since $u = -(B \otimes I_q) v$ from \eqref{eq:OutputCouplingAtEdges} and $(B \otimes I_q) Hw = (\Pi \otimes I_q) d$ from Lemma \ref{lem:Projection} in the Appendix, $v = Hw$ (at steady-state) implies $u_i + d_i = \frac{1}{N} \sum_{j \in \cN} d_j =: \bar d$ (see \eqref{eq:Dissipation}), where $v := [v_1; \cdots; v_E]$ and $w := [w_1; \cdots; w_N]$.
In other words, the net effect of disturbances on each system \eqref{eq:System} is the same for all $i = 1, \dots, N$ if $v = Hw$, which motivates us to consider controllers at the edges, able to generate the feedforward signal $v_g = H_g w$.

% THEOREM BEGIN: Controllers at Edges --------------------------------------------------------------------------------------------------------

\begin{thm} \label{thm:ContrAtEdges}
	The outputs of \eqref{eq:System} in closed-loop with the control
	\begin{subequations} \label{eq:ContrAtEdge}
	\begin{align}
			\dot\ze_g &= s(\ze_g) + H_g^\top \vrho_g, \label{eq:ContrAtEdge_IntModel} \\
			\dot\ka_g &= \del_g \vrho_g^\top \vrho_g, \qquad \qquad ~~ \del_g > 0, \label{eq:ContrAtEdge_Adaptation}\\
			v_g &= H_g \ze_g + \ka_g \vrho_g, \qquad \:\, g = 1, \dots, E  \label{eq:ContrAtEdge_Output}
	\end{align}
	\end{subequations}
	synchronize asymptotically if the corresponding closed-loop solution is bounded, where $s(\cdot)$ is defined by $s(w) := [s_1(w_1); \cdots; s_N(w_N)]$.
	Moreover, $\lim_{t \to \infty} \| y_i(t) - \bar y(t) \| = 0$ holds for all $i = 1, \dots, N$.
\end{thm}

% THEOREM END: Controllers at Edges ---------------------------------------------------------------------------------------------------------

% PROOF BEGIN: Controllers at Edges ---------------------------------------------------------------------------------------------------------

\begin{IEEEproof}
	Let us consider the function 
	\begin{align}
		W_1(x) = \frac{1}{2N} \sum_{i \in \cN} \sum_{j \in \cN} \Phi(x_i,x_j), \label{eq:W1}
	\end{align}
	which can be shown to satisfy $\ul\eta( \| \tl x \| ) \leq W_1(x) \leq \ol\eta( \| \tl x \| )$ for some class $\cK_\infty$ functions $\ul\eta$ and $\ol\eta$, again by Lemma \ref{lem:V1_Positivity} in the Appendix.
	Then, by using Lemma \ref{lem:CompactExpression} with $a_{ij}$ replaced by $1/N$, one obtains the time derivative of $W_1(x)$ as
	\begin{align*}
		\dot W_1(x) \leq \si y^\top (\Pi \otimes I_q) y + y^\top (\Pi \otimes I_q) u + y^\top (\Pi \otimes I_q) d.
	\end{align*}
	Let $\vrho := [\vrho_1; \cdots; \vrho_E]$, $\ze := [\ze_1; \cdots; \ze_E]$, $\ol H := \diag(H_1, \dots, H_E)$, and $\sfK := \diag(\ka_1, \dots, \ka_E)$.	
	Noting that $\Pi B = B$, $u = -(B \otimes I_q) v$, and $\vrho = (B^\top \otimes I_q) y$, one further has
	\begin{align*}
		\dot W_1(x) &\leq \si y^\top (\Pi \otimes I_q) y - y^\top (B \sfK B^\top \otimes I_q) y + y^\top (\Pi \otimes I_q) d - y^\top (B \otimes I_q) \ol H \ze.
	\end{align*}
	
	Let us now define $\tl\ze_g := \ze_g - w$ and $\tl\ka_g := \ka_g - \ka_\star$, where $\ka_\star \in \bR$ is a constant to be chosen shortly.
	Let us consider the function
	\begin{align*}
		W_2(\tl\ze, \tl\ka) = \frac{1}{2} \sum_{g = 1}^E \tl\ze_g^\top \tl\ze_g + \sum_{g = 1}^E \frac{\tl\ka_g^2}{2\del_g},
	\end{align*}
	where $\tl\ze := [\tl\ze_1; \cdots; \tl\ze_E]$ and $\tl\ka := [\tl\ka_1; \cdots; \tl\ka_E]$.
	Then, using the property \eqref{eq:Exosystem_Dissipatioin}, we obtain its time derivative as
	\begin{align*}
		\dot W_2(\tl\ze, \tl\ka) &= \sum_{g = 1}^E \Big\{ \tl\ze_g^\top \big( s(\ze_g) + H_g^\top \vrho_g - s(w) \big) 
				+ \tl\ka_g \vrho_g^\top \vrho_g \Big\} \\
		&\leq \tl\ze^\top \ol H^\top \vrho + \vrho^\top (\tl\sfK \otimes I_q) \vrho 
		= \tl\ze^\top \ol H^\top (B^\top \otimes I_q) y + y^\top (B \tl\sfK B^\top \otimes I_q) y, 
	\end{align*}
	in which $\tl\sfK := \diag(\tl\ka_1, \dots, \tl\ka_E) = \sfK - \ka_\star I_E$.
	
	Finally, let us consider the Lyapunov function $W(x, \tl\ze, \tl\ka) = W_1(x) + W_2(\tl\ze, \tl\ka)$ for the closed-loop system.
	Since $\tl\ze = \ze - (1_E \otimes w)$ by its construction and $(\Pi \otimes I_q) d = (BB^+ \otimes I_q) Rw = (B \otimes I_q) H w = (B \otimes I_q) \ol H (1_E \otimes w)$ by Lemma \ref{lem:Projection} in the Appendix, the derivative of $W(x, \tl\ze, \tl\ka)$ along the solutions of the system becomes
	\begin{align*}
		\dot W(x, \tl\ze, \tl\ka) &\leq \si y^\top (\Pi \otimes I_q) y - \ka_\star y^\top (L \otimes I_q) y \leq - (\ka_\star \lam_2 - \si) \tl y^\top \tl y,
	\end{align*}
	where $\lam_2$ and $\tl y$ are the ones in the proof of Theorem \ref{thm:ContrAtNodes}.
	If $\ka_\star > 0$ is chosen sufficiently large so that $\varepsilon := \ka_\star \lam_2 - \si > 0$, then the result follows from the same arguments in the proof of Theorem \ref{thm:ContrAtNodes}.
\end{IEEEproof}

% PROOF END: Controllers at Edges -----------------------------------------------------------------------------------------------------------

We note that the design \eqref{eq:ContrAtEdge} also possesses the flexibility discussed at the end of Section \ref{sec:ContrAtNodes}.
In particular, \eqref{eq:ContrAtEdge_Adaptation} with $v_g = \ka_g \vrho_g$ solves the problem in the absence of disturbances and does not require the strong coupling condition in \cite{Burger14}, while \eqref{eq:ContrAtEdge_IntModel} with $v_g = H_g \ze_g + \vrho_g$ achieves the asymptotic output synchronization of \eqref{eq:System} if $\lam_2 > \si$.

%%%%%%%%%%%%%%%%%%%%%%%%%%%%%%%%%%%%%%%%%%%%%%%%%%%%%%%%%%%%%%%%%%%%%%%%%%%%%%%%%%%%%%%%%%%%%%%%%%%%%%%%%%%%%%%%%%%%%%%%%%%%%%%%%%%%%%%%%%%%%%
\subsection{More on the proposed scheme} \label{sec:PhysicalLinks}
%%%%%%%%%%%%%%%%%%%%%%%%%%%%%%%%%%%%%%%%%%%%%%%%%%%%%%%%%%%%%%%%%%%%%%%%%%%%%%%%%%%%%%%%%%%%%%%%%%%%%%%%%%%%%%%%%%%%%%%%%%%%%%%%%%%%%%%%%%%%%%

% TikZ FIGURE BEGIN: Motivation of Physical Links -------------------------------------------------------------------------------------------

\begin{figure*}[tp]
	\centering%
	\resizebox{0.48\textwidth}{!}{%
	\subfloat[An electrical network consisting of two voltage-dependent current sources and one load.]{
	\begin{circuitikz}
		\ctikzset{bipoles/length=1cm}
		\draw	(0,0)	node[ground]{}
			to	[csI,i=$v_1$]				(0,1.7)
			to	[generic,l=$Y$]				(2.4,1.7)
			to	[generic,l_=$Y$,v^=$v_L$]	(2.4,0)		node[ground]{};
		\draw	(4.8,0)	node[ground]{}
			to	[csI,i_=$v_2$]				(4.8,1.7)
			to	[generic,l_=$Y$]			(2.4,1.7);
			
		\draw	(-0.65,1) node[anchor=south] {$+$};
		\draw	(-0.65,0.75) node[anchor=north] {$-$};
		\draw	(0,1.5) node[anchor=east] {$i_1$};
		
		\draw	(5.45,1) node[anchor=south] {$+$};
		\draw	(5.45,0.75) node[anchor=north] {$-$};
		\draw	(4.8,1.5) node[anchor=west] {$i_2$};

		\draw[thick,dotted,gray]	(4,2) -- (4.5,2.5);
		\draw[thick,dotted,gray]	(4.5,1.9) rectangle (7.2,3.2);
		
		\ctikzset{bipoles/length=.8cm}
		\draw (4.7,2.5)
			to	[short,o-]	(5.2,2.5) -- (5.2,2.8)
			to	[R]			(6.5,2.8) -- (6.5,2.5)
			to	[short,-o]	(7,2.5);
		\draw	(5.2,2.5)	--	(5.2,2.2)
			to	[L]			(6.5,2.2) -- (6.5,2.5);
	\end{circuitikz}\label{fig:ElecNetwork_Y}}
	}%
	\hfil
	\resizebox{0.48\textwidth}{!}{%
	\subfloat[An equivalent network of two nodes connected by a dynamic edge, obtained from the Kron reduction process.]{
	\begin{circuitikz}
		\ctikzset{bipoles/length=1cm}
		\draw	(0,0)	node[ground]{}
			to	[csI,i=$v_1$]				(0,1.7) -- (.8,1.7)
			to	[generic,l=$\cfrac{Y}{3}$]	(.8,0)	node[ground]{};
		\draw	(.8,1.7)
			to	[generic,l_=$Y/3$]			(5,1.7)
			to	[generic,l_=$\cfrac{Y}{3}$]	(5,0)	node[ground]{};
		\draw	(5.8,0)	node[ground]{}
			to	[csI,i_=$v_2$]				(5.8,1.7) -- (5,1.7);

		\draw	(-0.65,1) node[anchor=south] {$+$};
		\draw	(-0.65,0.75) node[anchor=north] {$-$};
		\draw	(0,1.5) node[anchor=east] {$i_1$};
		
		\draw	(6.45,1) node[anchor=south] {$+$};
		\draw	(6.45,0.75) node[anchor=north] {$-$};
		\draw	(5.8,1.5) node[anchor=west] {$i_2$};
			
		\draw[thick,red,dashed]		(-.9,-0.6) rectangle (1.5,2);
		\draw[thick,red,dashed]		(6.7,-0.6) rectangle (4.3,2);
		\draw[thick,blue,dashed]	(2.9,1.7) ellipse (1cm and 0.7cm);
		
		\node[red,rotate=90]	at (-1.1,.7) {Node 1};
		\node[red,rotate=-90]	at (6.9,.7) {Node 2};
		\draw[blue]				(2.9,2.4) node[anchor=south] {Dynamic edge};
	\end{circuitikz}\label{fig:ElecNetwork_Pi}}
	}%
	\caption{A motivating example of edges modeled by dynamical systems. $Y$ denotes the admittance of the electrical circuit in the dotted box and $v_L$ is the voltage across the load.}\label{fig:ElecNetwork}
\end{figure*}
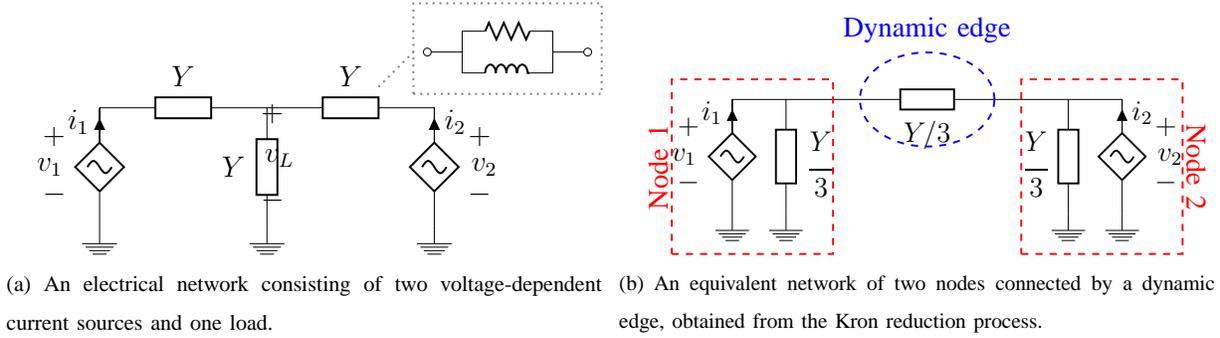

% TikZ FIGURE END: Motivation of Physical Links ---------------------------------------------------------------------------------------------

Our scheme in the previous subsections can be further applied to the case in which the dynamics at the edges are given and not free to design \cite{Torres15,Xiang14} such as electrical networks \cite{Dhople14,Johnson14}.
An example of those cases is illustrated in Fig. \ref{fig:ElecNetwork_Y}, where effective power transfer from the two sources to the load are aimed at.
Such power transfer will occur when the terminal voltages of the two sources are synchronized or, equivalently, when the two nodes coupled through the dynamic edge in Fig. \ref{fig:ElecNetwork_Pi} are synchronized.
Motivated by this example, let us consider a set of input strictly passive nonlinear systems, attached to the edges,
\begin{align*}
	\dot\eta_g = \psi_g(\eta_g, \vrho_g), \quad v_{g} = \vphi_g(\eta_g, \vrho_g), \quad g = 1,\dots, E,
\end{align*}
satisfying the dissipation inequality $\dot\Psi_g(\eta_g) \leq - \vrho_g^\top \vrho_g + v_{g}^\top \vrho_g$ for some positive definite $\Psi_g$, where $\psi_g$ and $\vphi_g$ are locally Lipschitz and continuous, respectively.

We now provide the following result.

% PROPOSITION BEGIN: Dynamic Edge, w/o Disturbances -----------------------------------------------------------------------------------------

\begin{prop}\label{prop:woDisturbances}
	Assume \eqref{eq:System} is disturbance-free, i.e., $\dot x_i = f(x_i,u_i)$, $y_i = h(x_i)$, and satisfies \eqref{eq:Dissipation} with $d_i = d_i' = 0$.
	Then, their outputs synchronize asymptotically if $u_i$ is given by \eqref{eq:OutputCouplingAtEdges}, the corresponding closed-loop solution is bounded, and the strong coupling condition $\lam_2 > \si$ is satisfied.
\end{prop}

% PROPOSITION END: Dynamic Edge, w/o Disturbances -------------------------------------------------------------------------------------------

% PROOF BEGIN: Dynamic Edge, w/o Disturbances -----------------------------------------------------------------------------------------------

\begin{IEEEproof}
	Let us consider the storage function $W(x,\eta) = W_1(x) + \sum_{g=1}^E \Psi_g(\eta_g)$, where $W_1(x)$ is in \eqref{eq:W1} and $\eta := [\eta_1; \cdots; \eta_E]$.
	Then, noting $u = -(B \otimes I_q) v$ and $\vrho = (B^\top \otimes I_q) y$, its time derivative is given by
	\begin{align*}
		\dot W(x,\eta)	&\leq \si y^\top (\Pi \otimes I_q) y + y^\top (\Pi \otimes I_q) u - \vrho^\top \vrho + v^\top \vrho \\
						&= \si y^\top (\Pi \otimes I_q) y - y^\top (BB^\top \otimes I_q) y \leq -(\lam_2 - \si) \tl y^\top \tl y.
	\end{align*}
	The rest can be proven by following the similar arguments in the proofs of Theorems \ref{thm:ContrAtNodes} and \ref{thm:ContrAtEdges}.
\end{IEEEproof}

% PROOF END: Dynamic Edge, w/o Disturbances -------------------------------------------------------------------------------------------------

The result of the proposition finds its application in problems of synchronization of, e.g., (a) electrical networks without shunt elements and with dynamic heterogeneous edges and (b) electrical networks with shunt elements in which their Kron-reduced networks\footnote{Refer to, e.g., \cite{Dorfler13} for the details on the Kron reduction process of electrical networks.} contain identical shunt elements with (possibly) heterogeneous dynamic edges and each node consisting of a source and its corresponding shunt element in the reduced networks is iOFP (refer to Fig. \ref{fig:ElecNetwork} as an example of this case).
Therefore, Proposition \ref{prop:woDisturbances} offers a complementary result and proof technique to \cite{Dhople14,Johnson14}.
We note however that in all cases, boundedness of the closed-loop solutions is indispensable for proving the synchronization.
Such boundedness and a class of systems including, e.g., Van der Pol oscillators will be discussed in Section \ref{sec:SolBoundedness}.
Oscillators which are not globally Lipschitz such as Van der Pol oscillators were not handled in \cite{Dhople14,Johnson14}.

When there are disturbances acting on the systems, the problem becomes more challenging.
However, at least on a complete graph, the outputs of the systems can be synchronized as follows.

% PROPOSITION BEGIN: Dynamic Edge, w/ Disturbances ------------------------------------------------------------------------------------------

\begin{prop}\label{prop:wDisturbances}
	Suppose that $a_{ij} = a$ for $i,j \in \cN$ and for some $a > 0$.
	Then, the outputs of the $N$ systems \eqref{eq:System} in closed-loop with the controllers
	\begin{align}
		\begin{split}
			\dot\xi_i &= s_i(\xi_i) - R_i^\top \rho_i, \qquad u_i = -R_i\xi_i - \sum_{g=1}^{E} b_{ig}v_g,
		\end{split} \qquad i = 1, 2, \cdots, N \label{eq:PhysicalLink_Contr_Disturbance}
	\end{align}
	synchronize asymptotically if $\lam_2 > \si$ and the corresponding closed-loop solution is bounded.
\end{prop}

% PROPOSITION END: Dynamic Edge, w/ Disturbances --------------------------------------------------------------------------------------------

% PROOF BEGIN: Dynamic Edge, w/ Disturbances ------------------------------------------------------------------------------------------------

\begin{IEEEproof}
	Let us consider the storage function $W(x,\eta,\tl\xi) = W_1(x) + \sum_{g=1}^E \Psi_g(\eta_g) + \sum_{i = 1}^N \tl\xi_i^\top \tl\xi_i/$ $(2aN)$, where $\tl\xi_i := \xi_i - w_i$ and $\tl\xi := [\tl\xi_1; \cdots; \tl\xi_N]$.
	Taking its time derivative and bearing in mind that $\rho = -aN (\Pi \otimes I_q) y$ and $u = -R\xi - (B \otimes I_q) v$, we get
	\begin{align*}
		\dot W(x,\eta,\tl\xi)	&\leq \si y^\top (\Pi \otimes I_q) y + y^\top (\Pi \otimes I_q) (u + d) - \vrho^\top \vrho + v^\top \vrho - \tl d^\top \rho / (aN) \\
								&= - y^\top \left(L - \si \Pi \otimes I_q\right) y \leq -(\lam_2 - \si) \tl y^\top \tl y,
	\end{align*}
	where $\tl d := R\xi - d$.
	Thus, the result again follows from the similar argument in the previous subsections.
\end{IEEEproof}

% PROOF END: Dynamic Edge, w/ Disturbances --------------------------------------------------------------------------------------------------

It is also possible to achieve the output synchronization without having the strong coupling conditions imposed in the previous propositions.
The idea is to consider the passive dynamical edges 
\begin{align*}
	\dot\eta_g = \psi_g(\eta_g, \vrho_g), \quad v_{1g} = \vphi_g(\eta_g, \vrho_g), \quad g = 1,\dots, E,
\end{align*}
satisfying $\dot\Psi_g(\eta_g) \leq v_{1g}^\top \vrho_g$ and to assign adaptive laws to the corresponding edges.
We note that such assignment may not be feasible in the case of, e.g., electrical networks.
However, the two corollaries given below are theoretically interesting and complement the results of the previous propositions in the sense that they provide adaptive variants of the propositions to relax strong coupling conditions.

% COROLLARY BEGIN: Dynamic Edge, Adaptation, w/o Disturbances -------------------------------------------------------------------------------

\begin{cor}\label{cor:woDisturbances_Adaptation}
	Assume the setup of Proposition \ref{prop:woDisturbances}.
	Then, the outputs of \eqref{eq:System} synchronize asymptotically by the control 
	\begin{align*}
		\dot{\ka}_g &=	\del_g \vrho_g^\top \vrho_g, \qquad v_{2g} = \ka_g \vrho_g, \qquad \del_g > 0, \qquad g = 1, \dots, E,\\
		u_i &=			-\sum_{g = 1}^E b_{ig} \left(v_{1g} + v_{2g}\right), \qquad\qquad\qquad\qquad\quad i = 1, \dots, N
	\end{align*}
	if the corresponding closed-loop solution is bounded.
\end{cor}

% COROLLARY END: Dynamic Edge, Adaptation, w/o Disturbances ---------------------------------------------------------------------------------

% PROOF BEGIN: Dynamic Edge, Adaptation, w/o Disturbances -----------------------------------------------------------------------------------

\begin{IEEEproof}
	Let us consider the function $W(x,\eta,\tl\ka) = W_1(x) + \sum_{g=1}^E \Psi_g(\eta_g) + \sum_{g=1}^E \tl\ka_g^2/(2\del_g)$, where $\tl \ka_g := \ka_g - \ka_\star$, $\ka_\star > 0$, and $\tl\ka := [\tl\ka_1; \cdots; \tl\ka_E]$.
	Let us define $v_i := [v_{i1}; \cdots; v_{iE}]$ with $i = 1,2$, and $v = v_1 + v_2$.
	Then, noting $u = -(B \otimes I_q) v$ and $v_2 = (\sfK \otimes I_q)\vrho = (\sfK B^\top \otimes I_q) y$ with $\sfK := \diag(\ka_1, \dots, \ka_E)$, one obtains the time derivative of $W$ as
	\begin{align*}
		\dot W(x,\eta,\tl\ka)	&\leq \si y^\top (\Pi \otimes I_q) y + y^\top (\Pi \otimes I_q) u + v_1^\top \vrho 
										+ \vrho^\top (\tl\sfK \otimes I_q) \vrho \\
								&= \si y^\top (\Pi \otimes I_q) y - y^\top (B \otimes I_q) v_2 + y^\top (B\tl\sfK B^\top \otimes I_q) y \\
								&= \si y^\top (\Pi \otimes I_q) y - \ka_\star y^\top (BB^\top \otimes I_q) y \leq -(\ka_\star \lam_2 - \si) \tl y^\top \tl y.
	\end{align*}
	The rest can be proven by following the similar arguments in the proofs of Theorems \ref{thm:ContrAtNodes} and \ref{thm:ContrAtEdges}.
\end{IEEEproof}

% PROOF END: Dynamic Edge, Adaptation, w/o Disturbances -------------------------------------------------------------------------------------

% COROLLARY BEGIN: Dynamic Edge, Adaptation, w/ Disturbances --------------------------------------------------------------------------------

\begin{cor}\label{cor:wDisturbances_Adaptation}
	Assume the setup of Proposition \ref{prop:wDisturbances}.
	Then, the outputs of the systems \eqref{eq:System} in closed-loop with 
	\begin{align*}
		\dot\xi_i &= s_i(\xi_i) - R_i^\top \rho_i, \qquad  \dot{\ka}_g =	\del_g \vrho_g^\top \vrho_g, \qquad \del_g > 0,\\
		u_i &= -R_i\xi_i - \sum_{g=1}^{E} b_{ig} \left(v_{1g} + v_{2g}\right), \qquad v_{2g} = \ka_g \vrho_g
	\end{align*}
	synchronize asymptotically whenever the solution is bounded.
\end{cor}

% COROLLARY END: Dynamic Edge, Adaptation, w/ Disturbances ----------------------------------------------------------------------------------

% PROOF BEGIN: Dynamic Edge, Adaptation, w/ Disturbances ------------------------------------------------------------------------------------

\begin{IEEEproof}
	Let us consider the function $W(x,\eta,\tl\xi,\tl\ka) = W_1(x) + \sum_{g=1}^E \Psi_g(\eta_g) + \sum_{i = 1}^N \tl\xi_i^\top \tl\xi_i/(2aN) + \sum_{g=1}^E \tl\ka_g^2/(2\del_g)$.
	Then, noting $u = - R\xi -(B \otimes I_q) v$, $\rho = -aN (\Pi \otimes I_q) y$, and $v_2 = (\sfK B^\top \otimes I_q) y$, we have
	\begin{align*}
		\dot W(x,\eta,\tl\ka)	&\leq \si y^\top (\Pi \otimes I_q) y + y^\top (\Pi \otimes I_q) (u + d) + v_1^\top \vrho - \tl d^\top \rho/(aN) + \vrho^\top (\tl\sfK \otimes I_q) \vrho \\
								&= \si y^\top (\Pi \otimes I_q) y - y^\top (B \otimes I_q) v_2 + y^\top (B\tl\sfK B^\top \otimes I_q) y \leq -(\ka_\star \lam_2 - \si) \tl y^\top \tl y.
	\end{align*}
	Thus, the result again follows from the similar arguments in the previous subsections.
\end{IEEEproof}

% PROOF END: Dynamic Edge, Adaptation, w/ Disturbances --------------------------------------------------------------------------------------

The extension of these results to the case of dynamics at the edges satisfying different dissipation inequalities from those assumed previously and to the case of graphs which are not complete, as well as the relaxation of the strong coupling condition by assigning adaptive laws to the nodes, is left for future investigation.

%%%%%%%%%%%%%%%%%%%%%%%%%%%%%%%%%%%%%%%%%%%%%%%%%%%%%%%%%%%%%%%%%%%%%%%%%%%%%%%%%%%%%%%%%%%%%%%%%%%%%%%%%%%%%%%%%%%%%%%%%%%%%%%%%%%%%%%%%%%%%%
%%%%%%%%%%%%%%%%%%%%%%%%%%%%%%%%%%%%%%%%%%%%%%%%%%%%%%%%%%%%%%%%%%%%%%%%%%%%%%%%%%%%%%%%%%%%%%%%%%%%%%%%%%%%%%%%%%%%%%%%%%%%%%%%%%%%%%%%%%%%%%
\section{Solution Boundedness} \label{sec:SolBoundedness}
%%%%%%%%%%%%%%%%%%%%%%%%%%%%%%%%%%%%%%%%%%%%%%%%%%%%%%%%%%%%%%%%%%%%%%%%%%%%%%%%%%%%%%%%%%%%%%%%%%%%%%%%%%%%%%%%%%%%%%%%%%%%%%%%%%%%%%%%%%%%%%
%%%%%%%%%%%%%%%%%%%%%%%%%%%%%%%%%%%%%%%%%%%%%%%%%%%%%%%%%%%%%%%%%%%%%%%%%%%%%%%%%%%%%%%%%%%%%%%%%%%%%%%%%%%%%%%%%%%%%%%%%%%%%%%%%%%%%%%%%%%%%%

In the previous section, the synchronization is guaranteed, provided that the solution of the closed-loop system is bounded.
We discuss in this section on what conditions of the open-loop system \eqref{eq:System} such boundedness is ensured under the control \eqref{eq:ContrAtNode} or \eqref{eq:ContrAtEdge}.
To do this, we further assume that the output map $h(\cdot)$ is globally Lipschitz and there is a compact set $\cX \subset \bR^n$, invariant for \eqref{eq:System} when $u_i \equiv 0$ and $d_i \equiv 0$, such that the system \eqref{eq:System} is input-to-state stable (ISS) with respect to $\hat u_i := [u_i; d_i]$ relative to $\cX$.
In other words, there are two functions $\be(\cdot,\cdot)$ and $\mu(\cdot)$ of class $\cK\cL$ and $\cK$, respectively, such that the solution of \eqref{eq:System} satisfies
\begin{align}
	\| x_i(t) \|_\cX \leq \max \left\{ \be( \| x_i(0) \|_\cX, t ), \mu \left( \sup_{0 \leq \tau \leq t} \| \hat u_i(\tau) \| \right) \right\}.
			\label{eq:ISS}
\end{align}
We refer the reader to \cite[Section III]{Isidori14} for some details of the condition \eqref{eq:ISS}.
A class of systems ensuring both conditions \eqref{eq:Dissipation} and \eqref{eq:ISS} will be discussed after presenting the following result.

% PROPOSITION BEGIN: Solution Boundedness ----------------------------------------------------------------------------------------------------

\begin{prop} \label{prop:SolBoundedness}
	Let the ISS property \eqref{eq:ISS} be satisfied and suppose either \eqref{eq:ContrAtNode} or \eqref{eq:ContrAtEdge} is applied to the system \eqref{eq:System}. 
	Then, the solutions of the closed-loop system are bounded and satisfy $\lim_{t \to \infty} \| y_i(t) - \bar y(t) \| = 0$, $i = 1, \dots, N$.
\end{prop}

% PROPOSITION END: Solution Boundedness ------------------------------------------------------------------------------------------------------

% PROOF BEGIN: Solution Boundedness ----------------------------------------------------------------------------------------------------------

\begin{IEEEproof}
	We prove the case for controllers placed at the nodes \eqref{eq:ContrAtNode}.
	Other cases in Sections \ref{sec:ContrAtEdges} and \ref{sec:PhysicalLinks} can be proven similarly.
	
	Let the variables $x$ and $\xi$ be the ones in the proof of Theorem \ref{thm:ContrAtNodes}, and define $k := [k_1; \cdots; k_N]$.
	Let $[0, T_u)$, $T_u < + \infty$ be the maximal time-interval, where the unique solution of the closed-loop system starting at $(x(0), \xi(0), k(0))$ exists.
	Then, \eqref{eq:ContrAtNode_dotV} holds on this interval.
	As a result, $\xi(t)$ and $k(t)$ are bounded for all $t \in [0, T_u)$ because $w(t)$ is bounded for $t \geq 0$ from \eqref{eq:Exosystem_Dissipatioin}.
	Note that $x_i(t) -  x_j(t)$, $i,j \in \cN$ are also bounded on the interval since $V_1(x)$ in \eqref{eq:V1} is positive definite with respect to $\tl x$.
	Together with the globally Lipschitz output map $h(x_i)$, this implies the boundedness of the control inputs $u_i(t)$, $i \in \cN$ on $[0, T_u)$.
	Therefore, the solution $x_i(t)$ is bounded from the ISS property \eqref{eq:ISS}, leading to the existence of a positive constant $M$ such that $\| [x(t); \xi(t); k(t)] \| \leq M$ for all $t \in [0, T_u)$.
	This means that the solution of the closed-loop system can be extended to $[0, T]$ for some $T > T_u$, which contradicts to the assumption that $[0, T_u)$ be the maximal interval of the solution existence.
	Thus, the result follows if we repeat the previous process with $T_u = +\infty$.
\end{IEEEproof}

% PROOF END: Solution Boundedness -----------------------------------------------------------------------------------------------------------

% TikZ FIGURE BEGIN: Lur'e-type Nonlienar Systems -------------------------------------------------------------------------------------------

\savebox\LureNonlin{%
	\begin{tikzpicture}
		\begin{axis}[
				width=2cm,	height=1.7cm,
				scale only axis,
				xmin=-1.5,	xmax=1.5,
				ymin=-1.1,	ymax=1.1,
				axis lines=middle,
				xtick=\empty,	ytick=\empty,
				ylabel=$\phi(y_i)$,
				every axis y label/.style={at={(current axis.above origin)},anchor=north east},
			]
			
			\addplot[
				domain=-1.3:1.3,
				smooth,
				forget plot,
			]{-x+x^3};
		\end{axis}
	\end{tikzpicture}
}

\begin{figure}[!t]
	\centering%
	\begin{tikzpicture}[>=stealth]
		\path	(0,0)	node[anchor=south] {$\bar u_i$}
				(1,0)	node[draw,circle] (Sum) {}
				(3.2,0)	node[draw,rectangle,text width=2cm,minimum height=1.1cm,text centered] (Sys) {\small Passive linear system $\Si$}
				(3.2,-1.8)	node[draw,rectangle,text width=2cm,text height=1.7cm] (Nonlin) {};
		\draw[->]	(-.8,0) -- (Sum.west);
		\draw[->]	(Sum.east) -- (Sys.west);
		\draw[->]	(Sys.east) -- (6.4,0);
		\draw[->]	(5.4,0) -- (5.4,-1.8) -- (Nonlin.east);
		\draw[->]	(Nonlin.west) -- (1,-1.8) -- (Sum.south);
		
		\draw	(-.1,0) node[anchor=north] {\scriptsize$(=u_i+d_i)$};
		\draw	(6,0) node[anchor=south] {$y_i$};
		\draw	(1,-0.5) node[anchor=west] {$-$};
		
		\node at (axis cs:3.28,-1.8) {\usebox\LureNonlin};		
	\end{tikzpicture}
	\caption{A Lur'e-type nonlinear system consisting of a passive linear system $\Si$ and a static nonlinearity $\phi(\cdot)$ in the feedback path.}
	\label{fig:LureSystem} 
\end{figure}
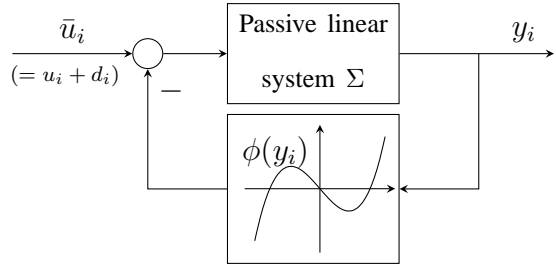

% TikZ FIGURE END: Lur'e-type Nonlienar Systems ---------------------------------------------------------------------------------------------

Theorem \ref{thm:ContrAtNodes} (or \ref{thm:ContrAtEdges}) combined with Proposition \ref{prop:SolBoundedness} requires the open-loop system \eqref{eq:System} to satisfy \eqref{eq:Dissipation} and \eqref{eq:ISS}, simultaneously.
We note that such systems indeed exist.
To see this, let us consider the Lur'e-type nonlinear systems (see Fig. \ref{fig:LureSystem} or \cite[Section 7]{Khalil02}) of the form:
\begin{align}
	\begin{split}
		\dot x_i &= A x_i + B \left( - \phi(y_i) + u_i + d_i \right), \qquad y_i = C x_i, 
	\end{split} \label{eq:LureSystem}
\end{align}
where $(C, A)$ is detectable and the linear part is passive, i.e., there is a matrix $P = P^\top > 0$ such that $A^\top P + PA \leq 0$ and $PB = C^\top$ hold.
The nonlinearity $\phi: \bR \rightarrow \bR$ is locally Lipschitz and satisfies that $\lim_{\tau \to +\infty} \phi(\tau) = +\infty$ and $\lim_{\tau \to -\infty} \phi(\tau) = -\infty$.
It is also assumed that there is $\tau_\star > 0$ such that $\phi(\tau)$ is monotonically increasing on both intervals $(-\infty, -\tau_\star]$ and $[\tau_\star, +\infty)$\footnote{An example of such nonlinearities is the function $\phi(\cdot)$ illustrated in Fig. \ref{fig:LureSystem}. We note that in contrast to \cite{Zhang14}, the nonlinearity $\phi$ considered here is neither incrementally passive nor incrementally sector bounded due to (possible) negative slope at the origin.}.
Then, we have the following.

% PROPOSITION BEGIN: iOFP and ISS of Lure System --------------------------------------------------------------------------------------------

\begin{prop} \label{prop:LureSystem_iOFP-ISS}
	The Lur'e feedback system \eqref{eq:LureSystem} satisfies both of the iOFP condition \eqref{eq:Dissipation} and the ISS property \eqref{eq:ISS}.
\end{prop}

% PROOF END: iOFP and ISS of Lure System ----------------------------------------------------------------------------------------------------

% PROOF BEGIN: iOFP and ISS of Lure System --------------------------------------------------------------------------------------------------

\begin{IEEEproof}
	The satisfaction of the ISS property \eqref{eq:ISS} follows from \cite[Thoerem 2]{Arcak02}. 
	To show the iOFP of \eqref{eq:LureSystem}, let us consider a storage function $\Phi(x_i, x_i') = \frac{1}{2} (x_i - x_i')^\top P (x_i - x_i')$ and its time derivative along the solutions $x_i$ and $x_i'$ of \eqref{eq:LureSystem} as
	\begin{align*}
		\dot\Phi(x_i,x_i')	&= (x_i - x_i')^\top P \left\{ Ax_i + B(-\phi(y_i) + u_i + d_i) - Ax_i' - B(-\phi(y_i') + u_i' + d_i') \right\} \\
							&\leq (y_i - y_i') \left( - \phi(y_i) + \phi(y_i') + u_i - u_i' + d_i - d_i' \right).
	\end{align*}
	This implies that \eqref{eq:LureSystem} is iOFP if there is $\si > 0$ satisfying $\pi(y_i, y_i') := - (y_i - y_i') \left( \phi(y_i) - \phi(y_i') \right) \leq \si (y_i - y_i')^2$. %, where $\pi(y_i, y_i') := - (y_i - y_i') \left( \phi(y_i) - \phi(y_i') \right)$. 
	The existence of such $\si$ can be shown as follows.
	\begin{enumerate}[i)]
		\item	$y_i, y_i' \in [-\tau_\star, \tau_\star]$: Since $\phi$ is locally Lipschitz, $\pi \leq \vpi (y_i - y_i')^2$ holds, where
				$\vpi > 0$ is a Lipschitz constant of $\phi$ on $[-\tau_\star, \tau_\star]$.
		\item	$y_i, y_i' \in [\tau_\star, +\infty)$: By the monotonicity of $\phi$ on $[\tau_\star, +\infty)$, it holds that $\pi \leq 0 
				\leq \vpi (y_i - y_i')^2$.
		\item	$y_i \in [-\tau_\star, \tau_\star]$ and $y_i' \in [\tau_\star, +\infty)$: From the Lipschitz property and monotonicity 
				of $\phi$, it follows that
				\begin{align*}
					\pi	&= -(y_i - y_i') \left(\phi(y_i) - \phi(\tau_\star)\right) 
								- (y_i - y_i') \left(\phi(\tau_\star) - \phi(y_i')\right) \\
						&\leq \vpi |y_i - y_i'| |y_i - \tau_\star| \leq \vpi (y_i - y_i')^2.
				\end{align*}
		\item	$y_i \in (-\infty, -\tau_\star]$ and $y_i' \in [\tau_\star, +\infty)$: Again, by the Lipschitzness and monotonicity 
				of $\phi$, we have
				\begin{align*}
					\pi	&= - (y_i - y_i') \left(\phi(y_i) - \phi(-\tau_\star)\right) - (y_i - y_i') \left(\phi(-\tau_\star) - \phi(\tau_\star)\right) - (y_i - y_i') \left(\phi(\tau_\star) - \phi(y_i')\right) \\
						&\leq \vpi |y_i - y_i'| |(-\tau_\star) - \tau_\star| \leq \vpi (y_i - y_i')^2.
				\end{align*}
	\end{enumerate}
	The remaining cases can be proven similarly.
	Therefore, $\pi(y_i, y_i') \leq \si (y_i - y_i')^2$ holds for all $y_i, y_i' \in \bR$ with $\si = \vpi > 0$.
\end{IEEEproof}

% PROOF END: iOFP and ISS of Lure System ----------------------------------------------------------------------------------------------------

% TikZ FIGURE BEGIN: Nonlinearity of Chua's Circuit -----------------------------------------------------------------------------------------

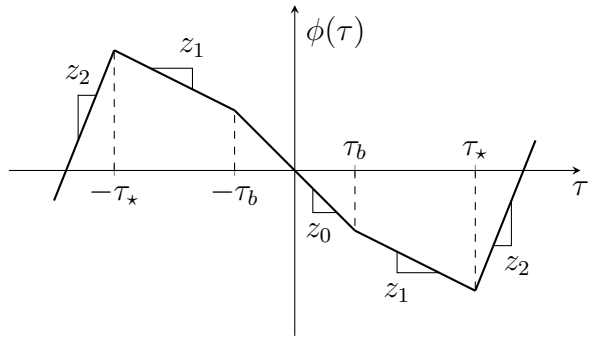
\begin{figure}[!t]
	\centering%
	\begin{tikzpicture}
		\begin{axis}[
				xmin=-9.5,	xmax=9.5,
				ymin=-5.5,	ymax=5.5,
				x=0.4cm,	y=0.4cm,
				axis lines=middle,
				xlabel=$\tau$,
				ylabel=$\phi(\tau)$,
				ytick=\empty,
				xtick={-6,-2,2,6},	xticklabels=\empty,
				every axis x label/.style={at={(current axis.right of origin)},anchor=north},
		        every axis y label/.style={at={(current axis.above origin)},anchor=north west},
			]
			
			\addplot[
				domain=-8:-6,
				thick
			]{2.5*x+19};
			
			\addplot[
				domain=-6:-2,
				thick
			]{-x/2+1};
			
			\addplot[
				domain=-2:2,
				thick
			]{-x};
			
			\addplot[
				domain=2:6,
				thick
			]{-x/2-1};
			
			\addplot[
				domain=6:8,
				thick
			]{2.5*x-19};
			
			\draw			(axis cs:-6,0) node[anchor=north] {$-\tau_\star$};
			\draw			(axis cs:-2,0) node[anchor=north] {$-\tau_{b}$};
			\draw			(axis cs:2,0) node[anchor=south] {$\tau_{b}$};		
			\draw			(axis cs:6,0) node[anchor=south] {$\tau_\star$};		
					
			\draw[dashed]	(axis cs:-6,0) -- (axis cs:-6,4);
			\draw[dashed]	(axis cs:-2,0) -- (axis cs:-2,2);
			\draw[dashed]	(axis cs:2,0) -- (axis cs:2,-2);
			\draw[dashed]	(axis cs:6,0) -- (axis cs:6,-4);
	
			\draw			(axis cs:-7.2,1) -- (axis cs:-7.2,2.5) -- (axis cs:-6.6,2.5);
			\draw			(axis cs:-3.4,2.7) -- (axis cs:-3.4,3.4) -- (axis cs:-4.8,3.4);
			\draw			(axis cs:0.6,-0.6) -- (axis cs:0.6,-1.4) -- (axis cs:1.4,-1.4);
			\draw			(axis cs:3.4,-2.7) -- (axis cs:3.4,-3.4) -- (axis cs:4.8,-3.4);
			\draw			(axis cs:7.2,-1) -- (axis cs:7.2,-2.5) -- (axis cs:6.6,-2.5);
			\draw			(axis cs:-7.2,2.5) node[anchor=south] {$z_2$};
			\draw			(axis cs:-3.4,3.4) node[anchor=south] {$z_1$};
			\draw			(axis cs:0.8,-1.4) node[anchor=north] {$z_0$};
			\draw			(axis cs:3.4,-3.4) node[anchor=north] {$z_1$};
			\draw			(axis cs:7.4,-2.5) node[anchor=north] {$z_2$};		
		\end{axis}
	\end{tikzpicture}
	\caption{The characteristic of the nonlinearity $\phi(\cdot)$ in Chua's circuit.}
	\label{fig:ChuaCircuit}
\end{figure}

% TikZ FIGURE END: Nonlinearity of Chua's Circuit -------------------------------------------------------------------------------------------

Examples of such Lur'e-type nonlinear systems include Van der Pol oscillators (see Section \ref{sec:Simulation} for the details) and the Li\'enard-type dead-zone oscillators (refer to \cite{Johnson14}).
Another example is the Chua's circuit \cite{Matsumoto85,Kennedy93} whose dimensionless form is given by
\begin{align}
	\begin{split}
		\dot x_{i,1} &= c_1 ( x_{i,2} - x_{i,1} - \phi(x_{i,1}) + u_i + d_i), \\
		\dot x_{i,2} &= x_{i,1} - x_{i,2} + x_{i,3}, \\
		\dot x_{i,3} &= -c_2 x_{i,2}, \qquad\qquad\qquad\qquad y_i = x_{i,1}, 
	\end{split} \label{eq:ChuaCircuit}
\end{align}
where $x_i := [x_{i,1}; x_{i,2}; x_{i,3}] \in \bR^3$, $c_1, c_2 > 0$, and the nonlinearity $\phi(\cdot)$ is shown in Fig. \ref{fig:ChuaCircuit}.
It is noted that the unforced system (i.e., \eqref{eq:ChuaCircuit} with $u_i \equiv 0$ and $d_i \equiv 0$) with particular selection of parameters $c_j$, $\tau_b$, $\tau_\star$, and $z_j$ has the double-scroll (chaotic) attractor (see, for instance, \cite[Fig. 6]{Matsumoto85} and \cite[Fig. 23]{Kennedy93}).
The Chua's circuit \eqref{eq:ChuaCircuit} is of the form \eqref{eq:LureSystem} and its linear part satisfies the passivity requirements $A^\top P + PA \leq 0$ and $PB = C^\top$ with $P = \diag(1/c_1, 1, 1/c_2)$.
Moreover, the piecewise linear function $\phi(\cdot)$ in Fig. \ref{fig:ChuaCircuit} guarantees all the required properties such as local Lipschitzness, monotonicity, and limiting behaviors at the infinity.
Therefore, by Proposition \ref{prop:LureSystem_iOFP-ISS}, the Chua's circuit \eqref{eq:ChuaCircuit} ensures both of the iOFP and ISS conditions.

%%%%%%%%%%%%%%%%%%%%%%%%%%%%%%%%%%%%%%%%%%%%%%%%%%%%%%%%%%%%%%%%%%%%%%%%%%%%%%%%%%%%%%%%%%%%%%%%%%%%%%%%%%%%%%%%%%%%%%%%%%%%%%%%%%%%%%%%%%%%%%
%%%%%%%%%%%%%%%%%%%%%%%%%%%%%%%%%%%%%%%%%%%%%%%%%%%%%%%%%%%%%%%%%%%%%%%%%%%%%%%%%%%%%%%%%%%%%%%%%%%%%%%%%%%%%%%%%%%%%%%%%%%%%%%%%%%%%%%%%%%%%%
\section{Computer Simulation: Van der Pol Oscillators} \label{sec:Simulation}
%%%%%%%%%%%%%%%%%%%%%%%%%%%%%%%%%%%%%%%%%%%%%%%%%%%%%%%%%%%%%%%%%%%%%%%%%%%%%%%%%%%%%%%%%%%%%%%%%%%%%%%%%%%%%%%%%%%%%%%%%%%%%%%%%%%%%%%%%%%%%%
%%%%%%%%%%%%%%%%%%%%%%%%%%%%%%%%%%%%%%%%%%%%%%%%%%%%%%%%%%%%%%%%%%%%%%%%%%%%%%%%%%%%%%%%%%%%%%%%%%%%%%%%%%%%%%%%%%%%%%%%%%%%%%%%%%%%%%%%%%%%%%

% FIGURE BEGIN: Simulation of Van der Pol Oscillators, III-A & III-B -------------------------------------------------------------------------

\begin{figure*}[!t]
	\centering%
	\subfloat[Trajectories of the states $x_{i,1}(t)$.]{\includegraphics[width=.45\columnwidth]{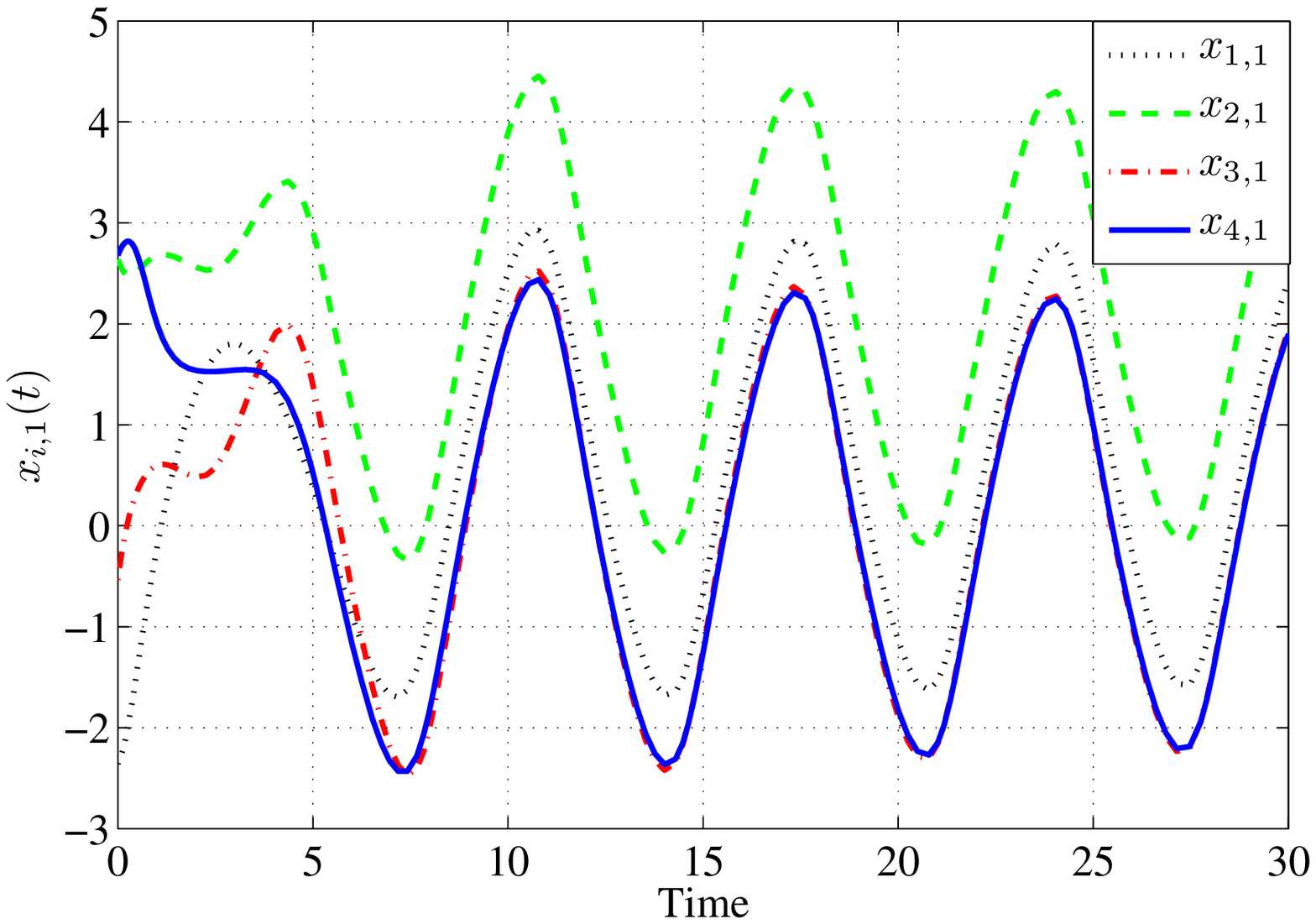}%
	\label{fig:Node_x1}}
	\hfil
	\subfloat[Trajectories of the states $x_{i,1}(t)$.]{\includegraphics[width=.45\columnwidth]{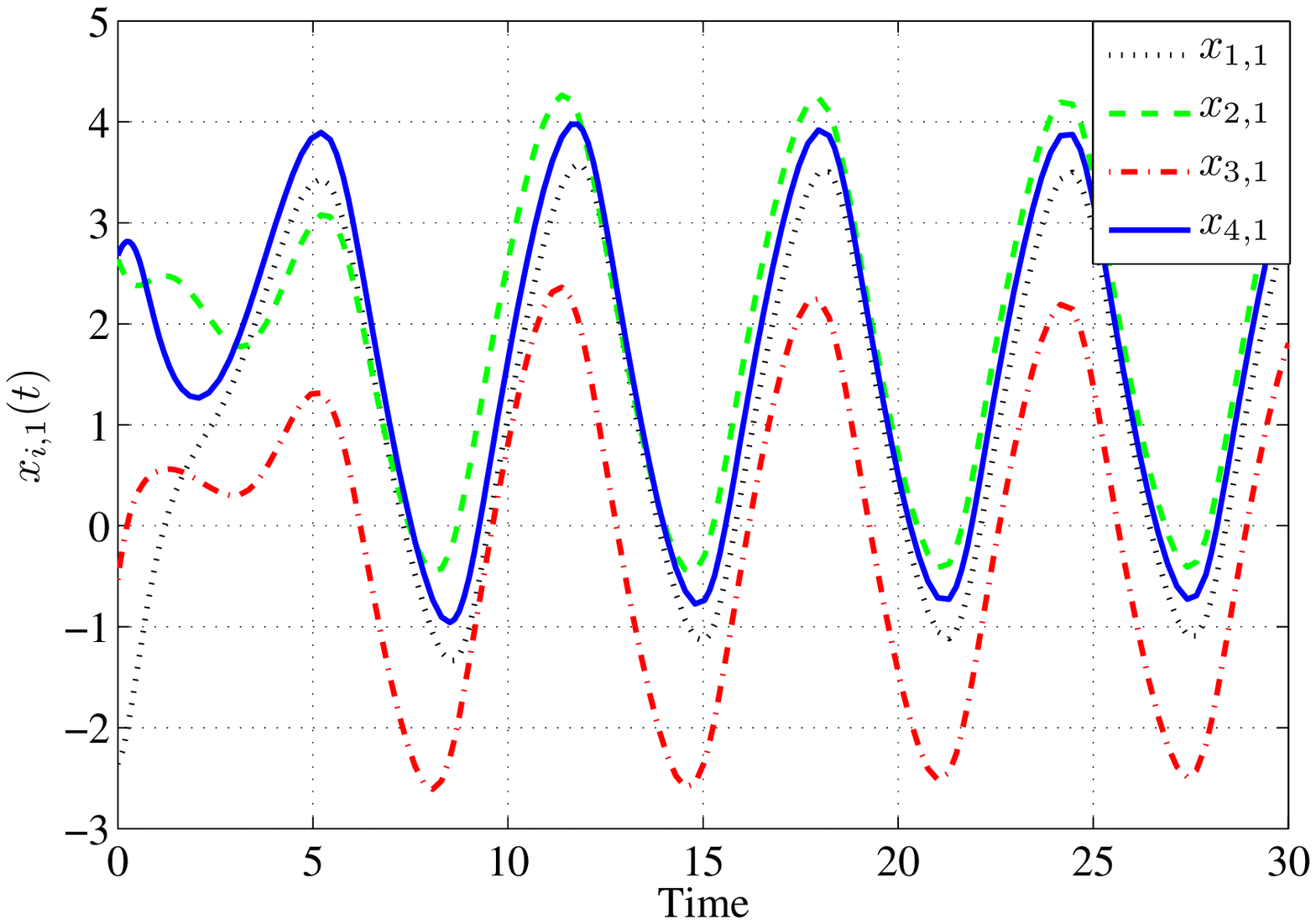}%
	\label{fig:Edge_x1}}\\%
	\subfloat[Trajectories of the outputs $y_i(t) = x_{i,2}(t)$.]{\includegraphics[width=.45\columnwidth]{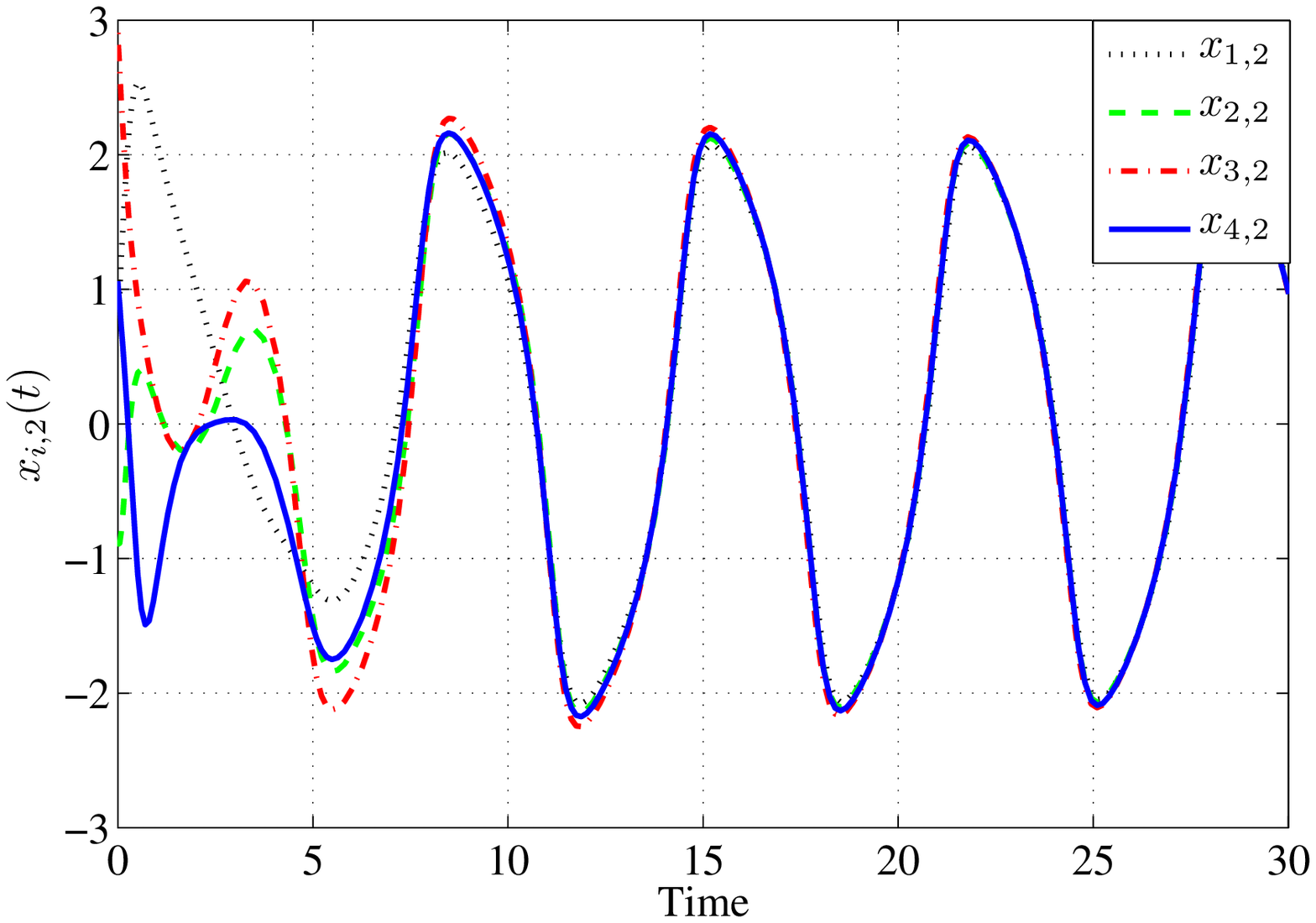}%
	\label{fig:Node_x2}}\hfil
	\subfloat[Trajectories of the outputs $y_i(t) = x_{i,2}(t)$.]{\includegraphics[width=.45\columnwidth]{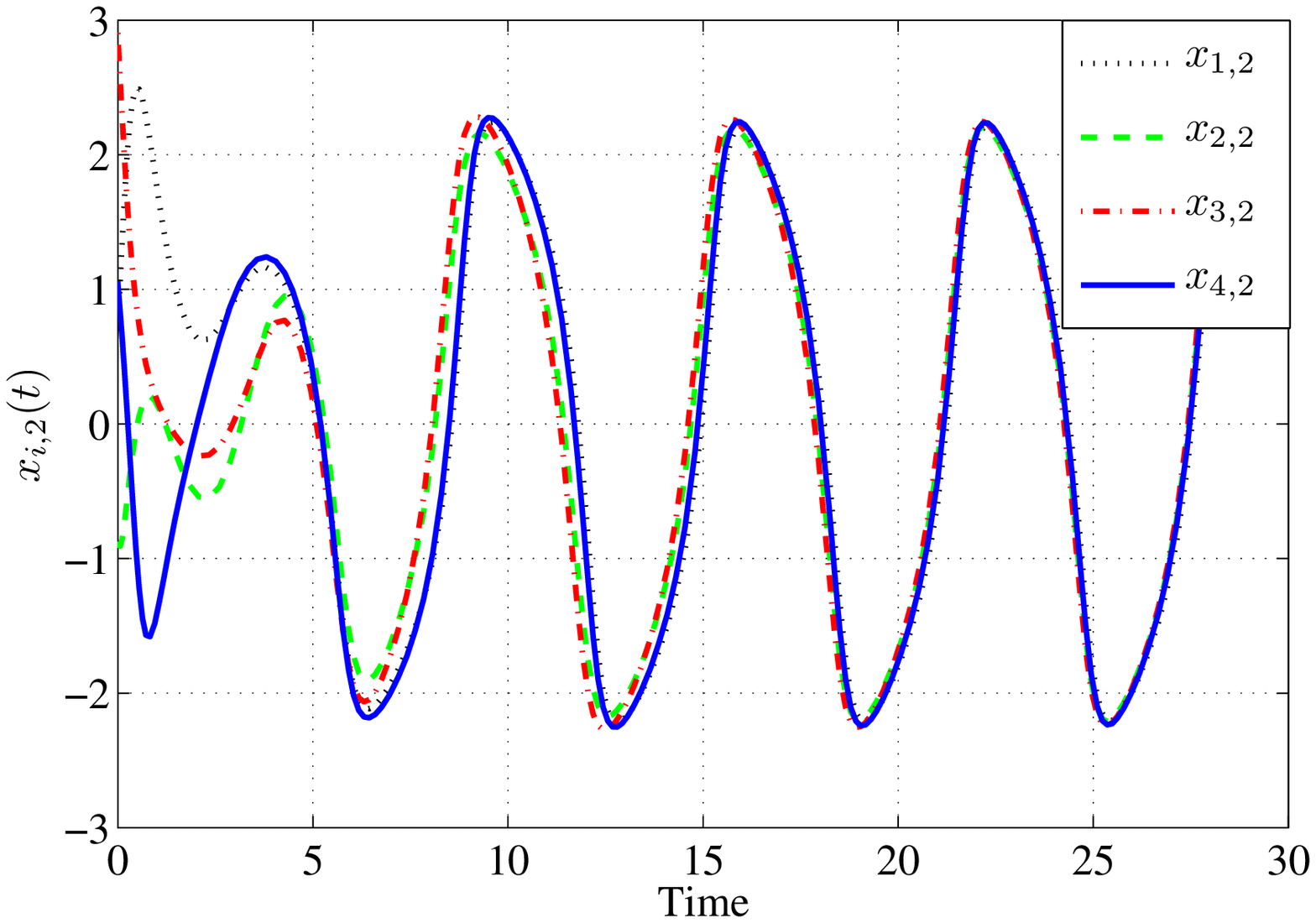}%
	\label{fig:Edge_x2}}
	\caption{Simulation results with controllers placed at the nodes (left) and at the edges (right).} \label{fig:Simulation}
\end{figure*}

% FIGURE END: Simulation of Van der Pol Oscillators, III-A & III-B -------------------------------------------------------------------------

Let us consider Van der Pol oscillators \cite[Example 2.9]{Khalil02} given by
\begin{align}
	\begin{split}
		\dot x_{i,1} &= x_{i,2}, \\
		\dot x_{i,2} &= - x_{i,1} + \nu (x_{i,2} - x_{i,2}^3/3) + u_i + d_i, \qquad i = 1, \dots, 4,\\
		y_i &= x_{i,2}, 
	\end{split} \label{eq:VanDerPol}
\end{align}
where $x_i := [x_{i,1}; x_{i,2}] \in \bR^2$ and $\nu \in \bR$ is a positive constant.
If we define $A := [0~ 1; -1 ~ 0]$, $B := [0; 1]$, $C := [0 ~ 1]$, and $\phi(y_i) := \nu (x_{i,2}^3/3 - x_{i,2})$, then \eqref{eq:VanDerPol} is of the form \eqref{eq:LureSystem} and satisfies all the required properties with $P = I_2$.
Thus, the Van der Pol oscillator \eqref{eq:VanDerPol} is iOFP and satisfies the ISS property given in \eqref{eq:ISS}.
Moreover, one can verify that $\si \geq \nu$.
We set $\nu = 1$.
Meanwhile, the disturbances $d_i$, $i = 1, \dots, 4$ in \eqref{eq:VanDerPol} are assumed to be generated by the exosystems (satisfying the requirement \eqref{eq:Exosystem_Dissipatioin})
\begin{alignat}{4}
	\dot w_i &= 0, &\quad d_i &= w_i, &\quad w_i &\in \bR, &\quad i &= 1,2, \nonumber\\
	\dot w_i &= \underbrace{\begin{bmatrix}
		0 & 1 \\
		-1 & 0
	\end{bmatrix} w_i}_{s_i(w_i)}, &\quad d_i &= \underbrace{\begin{bmatrix}
		1 & 0
	\end{bmatrix} w_i}_{R_i w_i}, &\quad w_i &\in \bR^2, &\quad i &= 3,4. \label{eq:VanDerPol_Disturbance}
\end{alignat}

In this setting, we first perform a set of simulations to demonstrate the results in Sections \ref{sec:ContrAtNodes} and \ref{sec:ContrAtEdges} with the interconnection structure $\cG$ characterized by its Laplacian $L = 0.09 \bar L$ and incidence matrix $B = 0.3 \bar B$, where
\begin{align*}
	\bar L = \begin{bmatrix}
		5	&	-1	&	0	&	-4 \\
		-1	&	14	&	-9	&	-4 \\
		0	&	-9	&	10	&	-1 \\
		-4	&	-4	&	-1	&	9
	\end{bmatrix}, \qquad \bar B = \begin{bmatrix}
		1	&	0	&	0	&	0	&	2 \\
		-1	&	3	&	2	&	0	&	0 \\
		0	&	-3	&	0	&	1	&	0\\
		0	&	0	&	-2	&	-1	&	-2
	\end{bmatrix}.
\end{align*}
Note that the strong coupling condition is not satisfied in this case since $\lam_2 = 0.4$ and $\si \geq 1$.
Figs. \ref{fig:Node_x1} and \ref{fig:Node_x2} show a simulation result with the controllers placed at the nodes \eqref{eq:ContrAtNode}, while Figs. \ref{fig:Edge_x1} and \ref{fig:Edge_x2} correspond to a simulation result with the controllers at the edges \eqref{eq:ContrAtEdge}.
The update gains $\ga_i$ and $\del_g$ are selected as $\ga_i = \del_g = 1$, $i = 1,\dots,4$, $g = 1, \dots, 5$.
All the elements of the initial conditions of the Van der Pol oscillators and exosystems are randomly chosen within the interval $[-3, 3]$, while those of controllers are set to all zeros.
From Figs. \ref{fig:Node_x2} and \ref{fig:Edge_x2}, it is seen that the outputs of the Van der Pol oscillators \eqref{eq:VanDerPol} synchronize asymptotically as expected.
Note however that the remaining states $x_{i,1}(t)$, $i = 1, \dots, 4$ do not synchronize since the proposed controllers guarantee the output synchronization only.

% FIGURE BEGIN: Simulation of Van der Pol Oscillators, III-C ---------------------------------------------------------------------------------

\begin{figure*}[!th]
	\centering%
	\subfloat[Trajectories of the states $x_{i,1}(t)$.]{\includegraphics[width=.45\columnwidth]{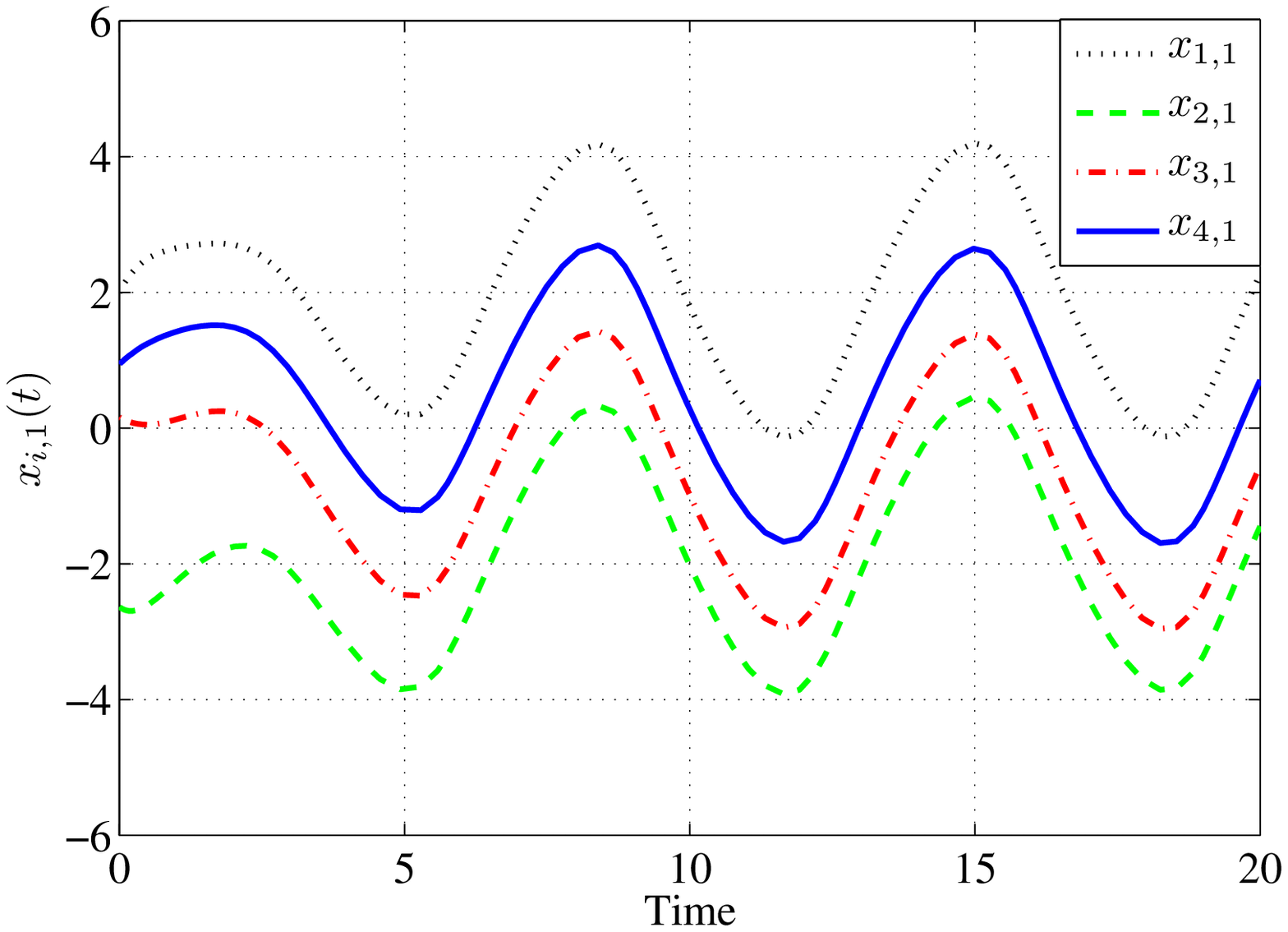}%
	\label{fig:woD_x1}}
	\hfil
	\subfloat[Trajectories of the states $x_{i,1}(t)$.]{\includegraphics[width=.45\columnwidth]{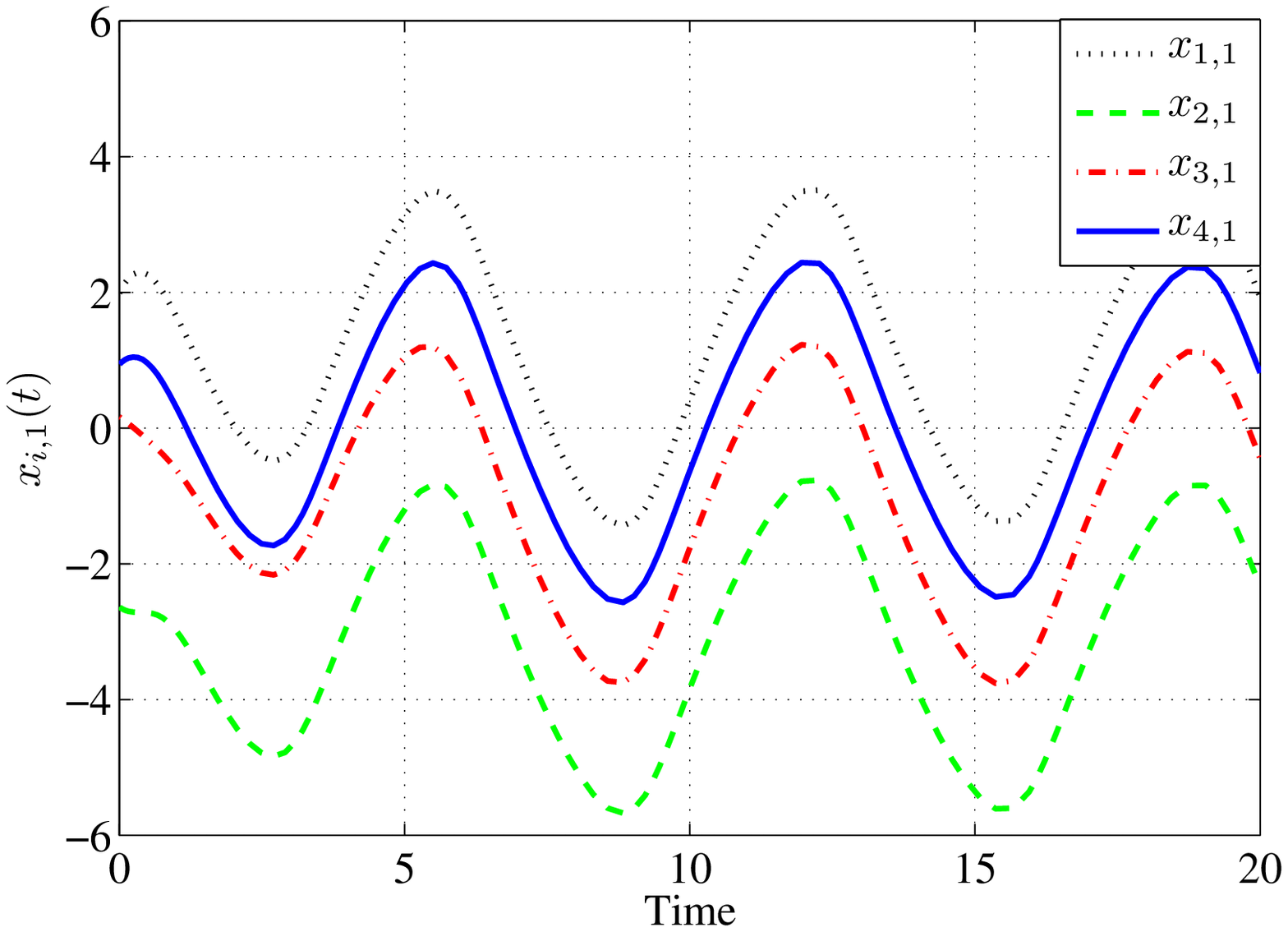}%
	\label{fig:wD_x1}}\\%
	\subfloat[Trajectories of the outputs $y_i(t) = x_{i,2}(t)$.]{\includegraphics[width=.45\columnwidth]{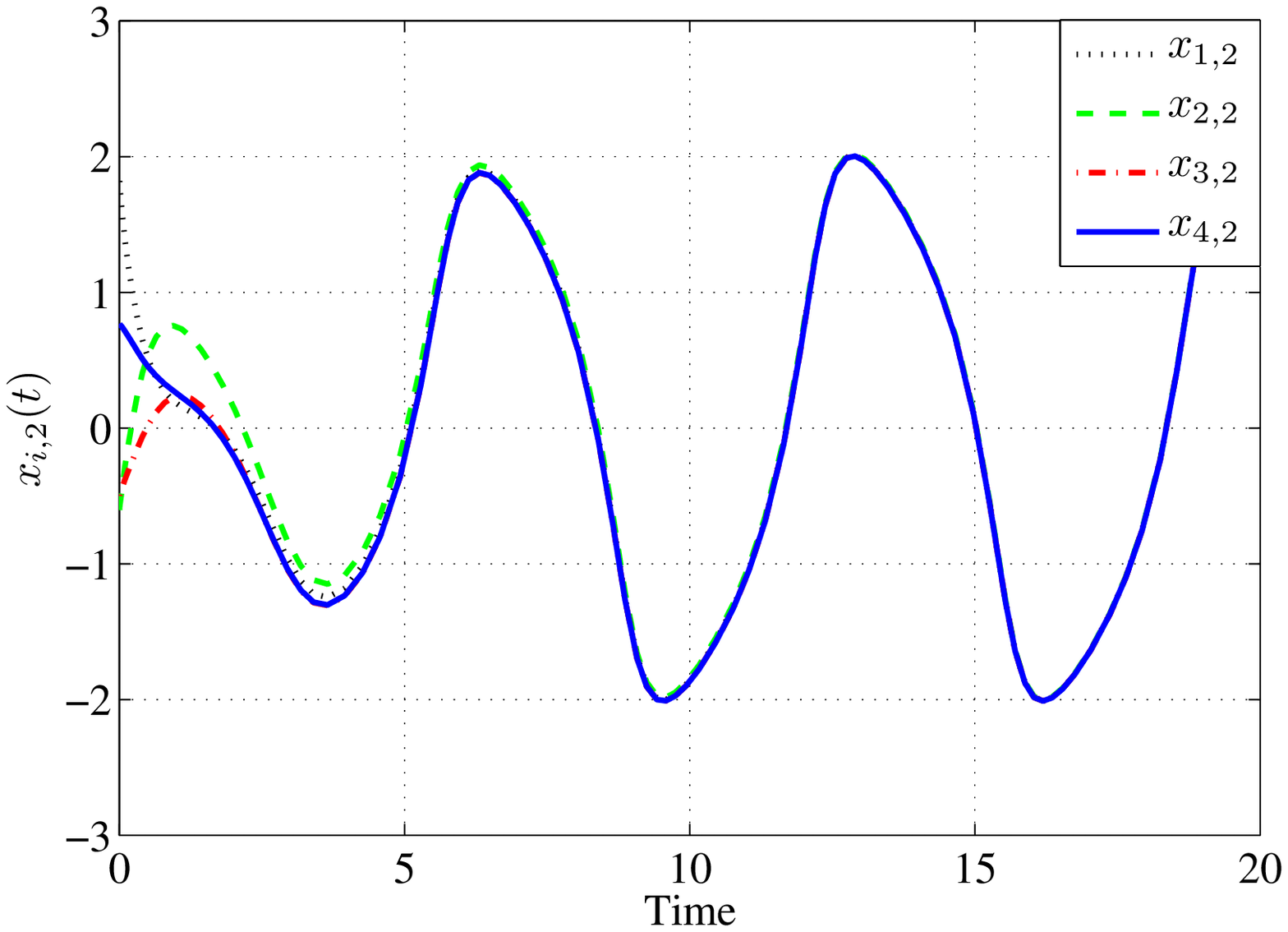}%
	\label{fig:woD_x2}}
	\hfil
	\subfloat[Trajectories of the outputs $y_i(t) = x_{i,2}(t)$.]{\includegraphics[width=.45\columnwidth]{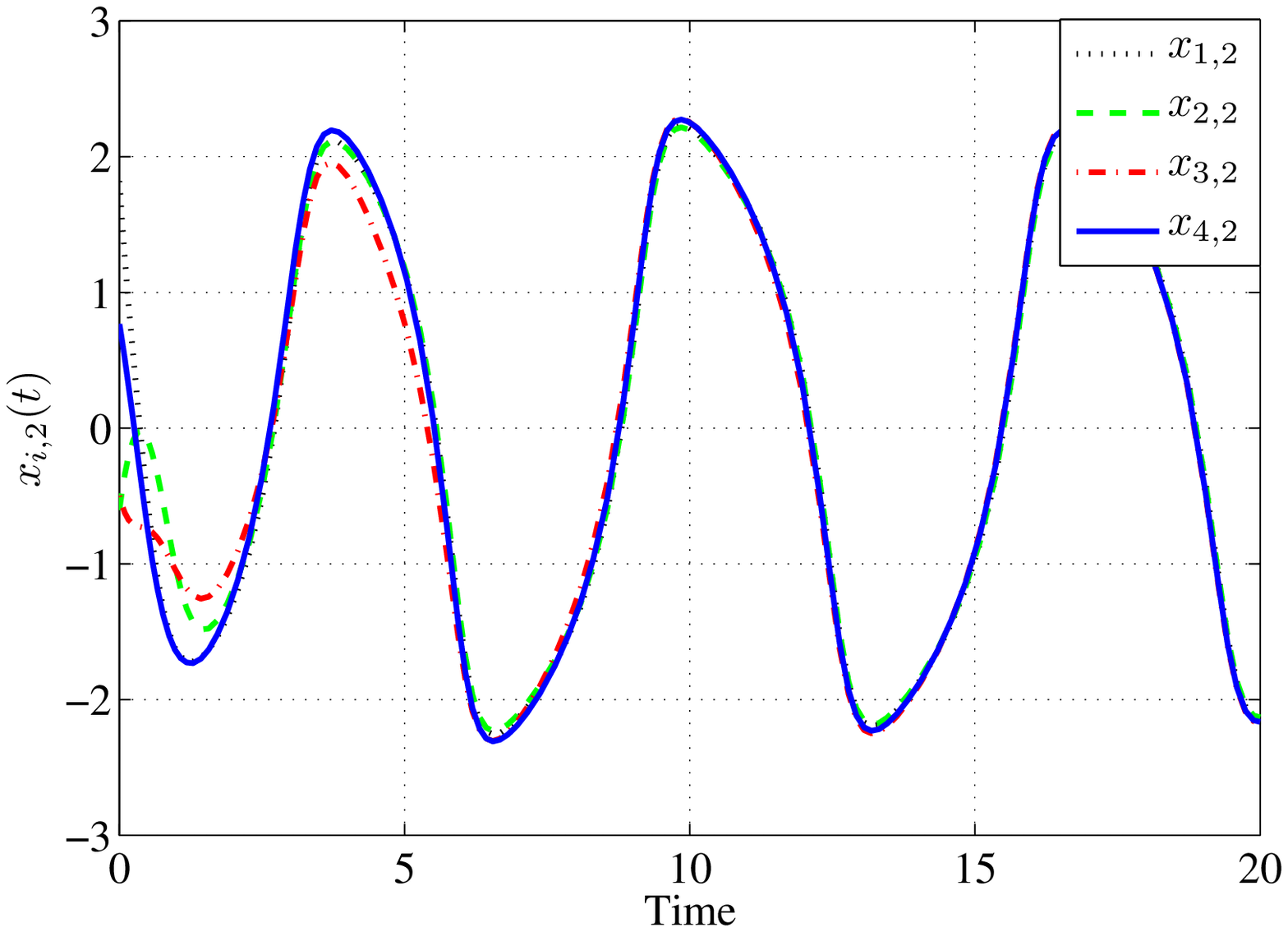}%
	\label{fig:wD_x2}}\\%
	\subfloat[Trajectories of the edge states $\eta_g(t)$.]{\includegraphics[width=.45\columnwidth]{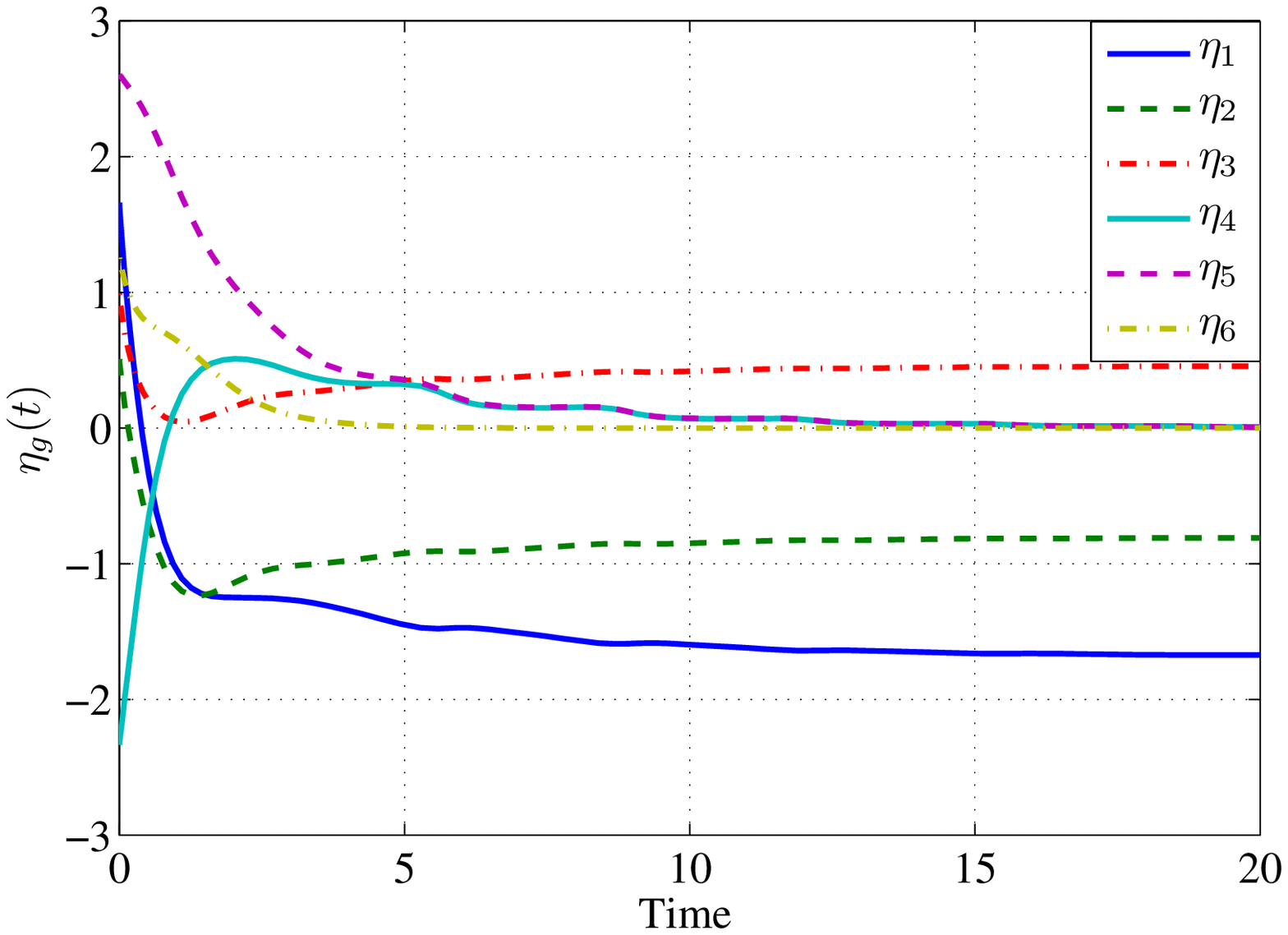}%
	\label{fig:woD_edge}}
	\hfil
	\subfloat[Trajectories of the edge states $\eta_g(t)$.]{\includegraphics[width=.45\columnwidth]{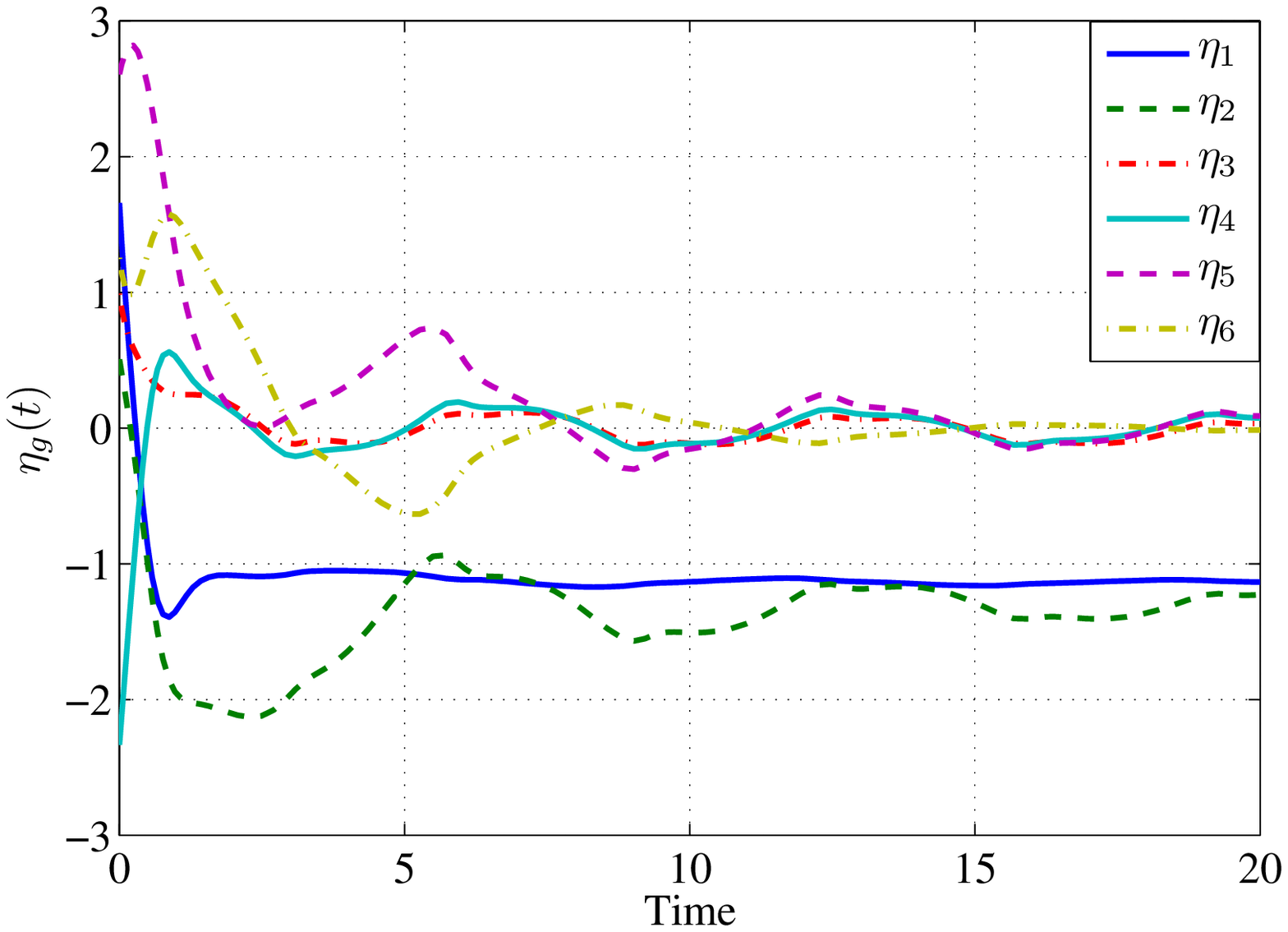}%
	\label{fig:wD_edge}}
	\caption{Simulation results in the presence of dynamical edges without disturbances (left) and with disturbances (right).} \label{fig:Simulation_PhysicalLink}
\end{figure*}

% FIGURE END: Simulation of Van der Pol Oscillators, III-C -----------------------------------------------------------------------------------

For the case of edges with dynamic systems attached as in Section \ref{sec:PhysicalLinks}, let us consider the systems \eqref{eq:VanDerPol} connected over the complete graph with $a_{ij} = 1$ for $i,j = 1, \dots, 4$, i.e., 
\begin{align*}
	L = 4\Pi, \qquad B = \begin{bmatrix}
		1	&	1	&	1	&	0	&	0	&	0 \\
		-1	&	0	&	0	&	1	&	1	&	0 \\
		0	&	-1	&	0	&	0	&	-1	&	1 \\
		0	&	0	&	-1	&	-1	&	0	&	-1
	\end{bmatrix}.
\end{align*}
In this case, the strong coupling condition is guaranteed since $\lam_2 = 4$ and the shortage of passivity, $\si$, can be taken as $\si = \nu = 1$.
The edges connected to the first Van der Pol oscillator are modeled by 
\begin{align*}
	\dot\eta_g = \vrho_g, \qquad\qquad\quad v_g = \eta_g + \vrho_g, \qquad g = 1, 2, 3,
\end{align*}
whereas the rest are given by the dynamical systems
\begin{align*}
	\dot\eta_g = -\eta_g + \vrho_g, \qquad v_g = \eta_g + \vrho_g \qquad g = 4, 5, 6.
\end{align*}
Note that with the storage function $\Psi(\eta_g) = \frac{1}{2} \eta_g^2$, both dynamics satisfy the dissipation inequality for the input strict passivity given in Section \ref{sec:PhysicalLinks}.
The left column of Fig. \ref{fig:Simulation_PhysicalLink} shows a simulation result with the controller \eqref{eq:OutputCouplingAtEdges} in the absence of disturbances, while the right column of Fig. \ref{fig:Simulation_PhysicalLink} shows a simulation result with the controller \eqref{eq:PhysicalLink_Contr_Disturbance} in the presence of the external disturbances generated by the systems \eqref{eq:VanDerPol_Disturbance}.
Each element of the initial conditions of the Van der Pol oscillators, exosystems, and dynamics at the edges are randomly chosen within the interval $[-3, 3]$, while those of controllers are set to all zeros.
As expected from Propositions \ref{prop:woDisturbances} and \ref{prop:wDisturbances}, the outputs of the Van der Pol oscillators \eqref{eq:VanDerPol} synchronize asymptotically. See the middle row of Fig. \ref{fig:Simulation_PhysicalLink}, i.e., Figs. \ref{fig:woD_x2} and \ref{fig:wD_x2}.

%%%%%%%%%%%%%%%%%%%%%%%%%%%%%%%%%%%%%%%%%%%%%%%%%%%%%%%%%%%%%%%%%%%%%%%%%%%%%%%%%%%%%%%%%%%%%%%%%%%%%%%%%%%%%%%%%%%%%%%%%%%%%%%%%%%%%%%%%%%%%%
%%%%%%%%%%%%%%%%%%%%%%%%%%%%%%%%%%%%%%%%%%%%%%%%%%%%%%%%%%%%%%%%%%%%%%%%%%%%%%%%%%%%%%%%%%%%%%%%%%%%%%%%%%%%%%%%%%%%%%%%%%%%%%%%%%%%%%%%%%%%%%
\section{Conclusion} \label{sec:Conclusion}
%%%%%%%%%%%%%%%%%%%%%%%%%%%%%%%%%%%%%%%%%%%%%%%%%%%%%%%%%%%%%%%%%%%%%%%%%%%%%%%%%%%%%%%%%%%%%%%%%%%%%%%%%%%%%%%%%%%%%%%%%%%%%%%%%%%%%%%%%%%%%%
%%%%%%%%%%%%%%%%%%%%%%%%%%%%%%%%%%%%%%%%%%%%%%%%%%%%%%%%%%%%%%%%%%%%%%%%%%%%%%%%%%%%%%%%%%%%%%%%%%%%%%%%%%%%%%%%%%%%%%%%%%%%%%%%%%%%%%%%%%%%%%

In this note, we have considered the asymptotic output synchronization problem of incrementally output-feedback passive nonlinear systems (defined over a connected undirected graph) in the presence of external disturbances.
Two different structures of distributed controllers are considered; one placed at the nodes and the other placed at the edges.
In both cases, the proposed controllers are shown to solve the problem if the solution of the corresponding closed-loop system is bounded. 
A pair of results has been also discussed to deal with the case in which the dynamics at the edges are given.
A class of incrementally output-feedback passive systems that guarantee the boundedness of the closed-loop solutions is then provided, by using the notion of input-to-state stability relative to a set.
The controllers are synthesized based on the adaptive control technique and the internal model principle and, as a consequence, do not require the so-called strong coupling condition.

%%%%%%%%%%%%%%%%%%%%%%%%%%%%%%%%%%%%%%%%%%%%%%%%%%%%%%%%%%%%%%%%%%%%%%%%%%%%%%%%%%%%%%%%%%%%%%%%%%%%%%%%%%%%%%%%%%%%%%%%%%%%%%%%%%%%%%%%%%%%%%
%%%%%%%%%%%%%%%%%%%%%%%%%%%%%%%%%%%%%%%%%%%%%%%%%%%%%%%%%%%%%%%%%%%%%%%%%%%%%%%%%%%%%%%%%%%%%%%%%%%%%%%%%%%%%%%%%%%%%%%%%%%%%%%%%%%%%%%%%%%%%%
\appendix
%%%%%%%%%%%%%%%%%%%%%%%%%%%%%%%%%%%%%%%%%%%%%%%%%%%%%%%%%%%%%%%%%%%%%%%%%%%%%%%%%%%%%%%%%%%%%%%%%%%%%%%%%%%%%%%%%%%%%%%%%%%%%%%%%%%%%%%%%%%%%%
%%%%%%%%%%%%%%%%%%%%%%%%%%%%%%%%%%%%%%%%%%%%%%%%%%%%%%%%%%%%%%%%%%%%%%%%%%%%%%%%%%%%%%%%%%%%%%%%%%%%%%%%%%%%%%%%%%%%%%%%%%%%%%%%%%%%%%%%%%%%%%

% LEMMA BEGIN: Positivity of V_1(x) ----------------------------------------------------------------------------------------------------------

\begin{lem} \label{lem:V1_Positivity}
	Suppose that $\cG$ is connected. Then, there are two class $\cK_\infty$ functions $\ul\eta$ and $\ol\eta$ that satisfy $\ul\eta( \| \tl x \| ) \leq V_1(x) \leq \ol\eta( \| \tl x \| )$, where $V_1(x)$ is in \eqref{eq:V1} and $\tl x = (\Pi \otimes I_n) x$.
\end{lem}

% LEMMA END: Positivity of V_1(x) ------------------------------------------------------------------------------------------------------------

% PROOF BEGIN: Positivity of V_1(x) ----------------------------------------------------------------------------------------------------------

\begin{IEEEproof}
	{\it (Existence of $\ol\eta$):} 
	Define $\tl x_i := x_i - \bar x$.
	Since
	\begin{align}
		\al(\tau_1 + \cdots + \tau_N) \leq \al(N \tau_1) + \cdots + \al(N \tau_N) \label{eq:ClassK_inf}
	\end{align}
	holds\footnote{Indeed, letting $\tau_i$ be one of the largest elements among $\tau_1, \dots, \tau_N$, we have that $\al(\tau_1 + \cdots + \tau_N) \leq \al(N \tau_i) \leq \al(N \tau_1) + \cdots + \al(N \tau_N)$.} for any class $\cK_\infty$ function $\al$, one has
	\begin{align*}
		V_1(x) &\leq \frac{1}{2} \sum_{i \in \cN} \sum_{j \in \cN} a_{ij} \ol\al( \| \tl x_i \| + \| \tl x_j \| ) \leq \frac{1}{2} \sum_{i \in \cN} \sum_{j \in \cN} a_{ij} \Big( \ol\al( 2\| \tl x_i \| ) + \ol\al( 2\| \tl x_j \| ) \Big) \\
		&= \sum_{i \in \cN} \vDel_i \ol\al( 2\| \tl x_i \| ) \leq \vDel_{\max} \sum_{i \in \cN} \ol\al( 2\| \tl x_i \| ), 
	\end{align*}
	where $\vDel_{\max} := \max_i \vDel_i > 0$.
	Defining $\tl X := [\| \tl x_1 \|; \cdots; \| \tl x_N \|] \in \bR^N$, the inequality further becomes
	\begin{align*}
		V_1(x) &\leq \vDel_{\max} \sum_{i \in \cN} \ol\al( 2\| \tl x_1 \| + \cdots + 2 \| \tl x_N \| ) = N \vDel_{\max} \ol\al( 2 \| \tl X \|_1 ) 
				\leq N \vDel_{\max} \ol\al( 2\sqrt{N} \| \tl X \| ).
	\end{align*}
	Observing $\| \tl X \| = \| \tl x \|$, set $\ol\eta(\tau) := N \vDel_{\max} \ol\al(2\sqrt{N} \tau)$.
	
	{\it (Existence of $\ul\eta$):}
	For $1 \leq i \leq N-1$ and $j > i$, let $\{ p_{ij,1}, p_{ij,2}, \dots, p_{ij,d_{ij}} \}$ be one of the shortest paths connecting the nodes $i$ and $j$. 
	Its length is $d_{ij} - 1$.
	The total length of those $N(N-1)/2$ paths satisfies
	\begin{align*}
		\sum_{i = 1}^{N-1} \sum_{j > i} (d_{ij} - 1) \leq \sum_{i = 1}^{N-1} \sum_{j > i} (N - 1) = \frac{N(N-1)^2}{2}.
	\end{align*}
	
	Define $a_{\min} := \min_{a_{ij} \neq 0} a_{ij} > 0$ and let $\cN_i$ be the set of neighbors of node $i$.
	Then, using \eqref{eq:ClassK_inf} and triangular inequality, and noting that $p_{ij,1} = i$ and $p_{ij,d_{ij}} = j$, we have
	\begin{align*}
		V_1(x) &\geq \frac{a_{\min}}{2} \sum_{i \in \cN} \sum_{j \in \cN_i} \ul\al(\| x_i - x_j \|) \geq \frac{a_{\min}}{N(N-1)^2} \sum_{i = 1}^{N-1} \sum_{j > i} \sum_{g = 1}^{d_{ij} - 1} \ul\al(\| x_{p_{ij,g}} - x_{p_{ij,g+1}} \|) \\
		&\geq \frac{a_{\min}}{N(N-1)^2} \sum_{i = 1}^{N-1} \sum_{j > i} \ul\al\left( 
				\frac{\sum_{g = 1}^{d_{ij} - 1} \| x_{p_{ij,g}} - x_{p_{ij,g+1}} \| }{d_{ij} - 1} \right) \\
		&\geq \frac{a_{\min}}{N(N-1)^2} \sum_{i = 1}^{N-1} \sum_{j > i} \ul\al\left( \frac{\| x_i - x_j \|}{N - 1} \right) 
		= \frac{a_{\min}}{2N(N-1)^2} \sum_{i = 1}^N \sum_{j = 1}^N \ul\al\left( \frac{\| x_i - x_j \|}{N - 1} \right).
	\end{align*}
	By using \eqref{eq:ClassK_inf} and triangular inequality again, one finally has
	\begin{align*}
		V_1(x) &\geq \frac{a_{\min}}{2N(N-1)^2} \ul\al\left( \frac{1}{N(N - 1)^2} \sum_{i = 1}^N \sum_{j = 1}^N \| x_i - x_j \| \right) \\
		&\geq \frac{a_{\min}}{2N(N-1)^2} \ul\al\left( \frac{1}{(N - 1)^2} \sum_{i = 1}^N \| \tl x_i \| \right) 
		\geq \frac{a_{\min}}{2N(N-1)^2} \ul\al\left( \frac{\| \tl X \|}{(N - 1)^2} \right).
	\end{align*}
	Set $\ul\eta(\tau) := a_{\min} \ul\al(\tau/(N-1)^2) / \{2N(N-1)^2\}$.
\end{IEEEproof}

% PROOF END: Positivity of V_1(x) ----------------------------------------------------------------------------------------------------------

% LEMMA BEGIN: Compact Expression ----------------------------------------------------------------------------------------------------------

\begin{lem} \label{lem:CompactExpression}
	Let $\theta_i, \vartheta_i \in \bR^q$ for $i = 1, \dots, N$. Then, the following holds.
	\begin{align*}
		\frac{1}{2} \sum_{i \in \cN} \sum_{j \in \cN} a_{ij} (\theta_i - \theta_j)^\top (\vartheta_i - \vartheta_j) 
				= \theta^\top (L \otimes I_q) \vartheta,
	\end{align*}
	where $\theta := [\theta_1; \cdots; \theta_N]$ and $\vartheta := [\vartheta_1; \cdots; \vartheta_N]$.
\end{lem}

% LEMMA END: Compact Expression -----------------------------------------------------------------------------------------------------------

% PROOF BEGIN: Compact Expression ---------------------------------------------------------------------------------------------------------

\begin{IEEEproof}
	The following computation proves the lemma.
	\begin{align*}
		\frac{1}{2} \sum_{i \in \cN} \sum_{j \in \cN} a_{ij} (\theta_i - \theta_j)^\top (\vartheta_i - \vartheta_j) &= \sum_{i \in \cN} \vDel_i \theta_i^\top \vartheta_i - \frac{1}{2} \sum_{i \in \cN} \Big( \theta_i^\top (A_i \otimes I_q) \vartheta 
				+ \theta^\top (A_i^\top \otimes I_q) \vartheta_i \Big) \\
		&= \theta^\top (\vDel \otimes I_q) \vartheta - \theta^\top (A \otimes I_q) \vartheta
				= \theta^\top (L \otimes I_q) \vartheta,
	\end{align*}
	where $A_i$ is the $i$-th row of symmetric adjacency matrix $A$.
\end{IEEEproof}

% PROOF END: Compact Expression -----------------------------------------------------------------------------------------------------------

% LEMMA BEGIN: Moore-Penrose Pseudoinverse ------------------------------------------------------------------------------------------------

\begin{lem} \label{lem:Projection}
	Let $\cG$ be a connected undirected graph.
	Then, $BB^+ = LL^+ = \Pi$ holds.
\end{lem}

% LEMMA END: Moore-Penrose Pseudoinverse --------------------------------------------------------------------------------------------------

% PROOF BEGIN: Moore-Penrose Pseudoinverse ------------------------------------------------------------------------------------------------

\begin{IEEEproof}
	Since $B^+ = B^\top (BB^\top)^+$ by \cite[p. 49]{Ben-Israel03} and $L = BB^\top$, we have $BB^+ = LL^+$.
	On the other hand, by \cite[p. 60, Corollary 7]{Ben-Israel03}, $LL^+$ is the unique (orthogonal) projector on ${\rm im}(L)$ along ${\rm ker}(L^+) = {\rm ker}(L^\top) = {\rm ker}(L)$, where ${\rm im}(L)$ and ${\rm ker}(L)$ denote the image and kernel of $L$, respectively.
	Therefore, $LL^+ = \Pi$ because $\Pi$ is also the projector on $1_N^\bot = {\rm im}(L)$ along ${\rm span}(1_N) = {\rm ker}(L^+)$, where $1_N^\bot$ is the orthogonal complement of $1_N$.
%	Since $L$ and $\Pi$ commute (i.e., $L \Pi = \Pi L$), they are simultaneously diagonalizable \cite[Fact 8.16.1]{Bernstein09}.
%	In particular, there is a unitary matrix $U$ such that $L = U\Lam U^*$ and $\Pi = U \diag(0, I_{N-1}) U^*$, where $*$ denotes the conjugate transpose and $\Lam$ is the one in the proof of Theorem \ref{thm:ContrAtNodes}.
%	Let $\Lam^+ := \diag(0, 1/\lam_2, \dots, 1/\lam_N)$.
%	Then, $L^+ =$ $ U \Lam^+ U^*$ by \cite[p. 49]{Ben-Israel03} and hence, $L L^+ = U \Lam \Lam^+ U^* = \Pi$.
\end{IEEEproof}

% PROOF END: Moore-Penrose Pseudoinverse --------------------------------------------------------------------------------------------------

%% use section* for acknowledgement
%\section*{Acknowledgment}
%The authors would like to thank...

% Bibliography ----------------------------------------------------------------------------------------------------------------------------
\bibliographystyle{IEEEtran}
\bibliography{hkkim} %C:/Users/HKKIM/Documents/Work/02.WinEdt/hkkim}

% Generated by IEEEtran.bst, version: 1.12 (2007/01/11)
\begin{thebibliography}{10}
\providecommand{\url}[1]{#1}
\csname url@samestyle\endcsname
\providecommand{\newblock}{\relax}
\providecommand{\bibinfo}[2]{#2}
\providecommand{\BIBentrySTDinterwordspacing}{\spaceskip=0pt\relax}
\providecommand{\BIBentryALTinterwordstretchfactor}{4}
\providecommand{\BIBentryALTinterwordspacing}{\spaceskip=\fontdimen2\font plus
\BIBentryALTinterwordstretchfactor\fontdimen3\font minus
  \fontdimen4\font\relax}
\providecommand{\BIBforeignlanguage}[2]{{%
\expandafter\ifx\csname l@#1\endcsname\relax
\typeout{** WARNING: IEEEtran.bst: No hyphenation pattern has been}%
\typeout{** loaded for the language `#1'. Using the pattern for}%
\typeout{** the default language instead.}%
\else
\language=\csname l@#1\endcsname
\fi
#2}}
\providecommand{\BIBdecl}{\relax}
\BIBdecl

\bibitem{Arcak07}
M.~Arcak, ``Passivity as a design tool for group coordination,'' \emph{IEEE
  Transactions on Automatic Control}, vol.~52, no.~8, pp. 1380--1390, Aug.
  2007.

\bibitem{Stan07a}
G.-B. Stan and R.~Sepulchre, ``Analysis of interconnected oscillators by
  dissipativity theory,'' \emph{IEEE Transactions on Automatic Control},
  vol.~52, no.~2, pp. 256--270, Feb. 2007.

\bibitem{Pogromsky01}
A.~Pogromsky and H.~Nijmeijer, ``Cooperative oscillatory behavior of mutually
  coupled dynamical systems,'' \emph{IEEE Transactions on Circuits and
  Systems---I: Fundamental Theory and Applications}, vol.~48, no.~2, pp.
  152--162, Feb. 2001.

\bibitem{Scardovi10}
L.~Scardovi, M.~Arcak, and E.~D. Sontag, ``Synchronization of interconnected
  systems with applications to biochemical networks: {A}n input-output
  approach,'' \emph{IEEE Transactions on Automatic Control}, vol.~55, no.~6,
  pp. 1367--1379, Jan. 2010.

\bibitem{Shafi14}
S.~Y. Shafi and M.~Arcak, ``An adaptive algorithm for synchronization in
  diffusively-coupled systems,'' in \emph{Proceedings of the 2014 American
  Control Conference}, 2014, pp. 2220--2225.

\bibitem{Bai13}
H.~Bai and S.~Y. Shafi, ``Output synchronization of nonlinear systems under
  input disturbances,'' arXiv:1312.6421v1 [cs.SY], 2013, {A}vailable at
  \url{http://arxiv.org/abs/1312.6421}.

\bibitem{Claudio14}
C.~{De Persis} and B.~Jayawardhana, ``On the internal model principle in the
  coordination of nonlinear systems,'' \emph{IEEE Transactions on Control of
  Network Systems}, vol.~1, no.~3, pp. 272--282, Sep. 2014.

\bibitem{Burger15}
M.~B\"{u}rger and C.~{De Persis}, ``Dynamic coupling design for nonlinear
  output agreement and time-varying flow control,'' \emph{Automatica}, vol.~51,
  no.~1, pp. 210--222, Jan. 2015.

\bibitem{Burger14}
------, ``Further result about dynamic coupling for nonlinear output
  agreement,'' in \emph{Proceedings of the 53rd IEEE Conference on Decision and
  Control}, 2014, pp. 1353--1358.

\bibitem{Torres15}
L.~A.~B. T\^{o}rres, J.~P. Hespanha, and J.~Moehlis, ``Synchronization of
  identical oscillators coupled through a symmetric network with dynamics: {A}
  constructive approach with applications to parallel operation of inverters,''
  {S}ubmitted, Available at \url{http://www.ece.ucsb.edu/~hespanha/published}.

\bibitem{Wei13}
J.~Wei and A.~J. {van der Schaft}, ``Load balancing of dynamical distribution
  networks with flow constraints and unknown in/outflows,'' \emph{Systems \&
  Control Letters}, vol.~62, no.~11, pp. 1001--1008, Nov. 2013.

\bibitem{Bai11}
H.~Bai, M.~Arcak, and J.~Wen, \emph{Cooperative control design: {A} systematic,
  passivity-based approach}.\hskip 1em plus 0.5em minus 0.4em\relax Springer,
  2011, vol. 89.

\bibitem{Burger14a}
M.~B\"{u}rger, C.~{De Persis}, and S.~Trip, ``An internal model approach to
  (optimal) frequency regulation in power grids,'' in \emph{Proceedings of the
  21th International Symposium on Mathematical Theory of Networks and Systems},
  2014, pp. 577--583.

\bibitem{Burger14b}
M.~B\"{u}rger, C.~{De Persis}, and F.~Allg\"{o}wer, ``Dynamic pricing control
  for constrained distribution networks with storage,'' 2014, {\it {IEEE}
  Transactions on Control of Network Systems}, in press, DOI:
  \url{10.1109/TCNS.2014.2367572}.

\bibitem{Khalil02}
H.~K. Khalil, \emph{Nonlinear systems}, 3rd~ed.\hskip 1em plus 0.5em minus
  0.4em\relax Prentice Hall, 2002.

\bibitem{Byrnes97}
C.~I. Byrnes, F.~D. Priscoli, and A.~Isidori, \emph{Output regulation of
  uncertain nonlinear systems}.\hskip 1em plus 0.5em minus 0.4em\relax MA:
  Birkha\"{u}ser, Boston, 1997.

\bibitem{Xiang14}
J.~Xiang, Y.~Li, and D.~J. Hill, ``Cooperative output regulation of multi-agent
  systems coupled by dynamic edges,'' in \emph{Proceedings of the 19th IFAC
  World Congress}, 2014, pp. 1813--1818.

\bibitem{Godsil04}
C.~D. Godsil and G.~Royle, \emph{Algebraic graph theory}.\hskip 1em plus 0.5em
  minus 0.4em\relax Springer, 2001, {G}raduate {T}exts in {M}athematics, vol.
  207.

\bibitem{Ben-Israel03}
A.~Ben-Israel and T.~N.~E. Greville, \emph{Generalized inverses: Theory and
  applications}, 2nd~ed.\hskip 1em plus 0.5em minus 0.4em\relax Springer, 2003.

\bibitem{Tao97}
G.~Tao, ``A simple alternative to the {B}arb\u{a}lat lemma,'' \emph{IEEE
  Transactions on Automatic Control}, vol.~42, no.~5, p. 698, May 1997.

\bibitem{Li13b}
Z.~Li, W.~Ren, X.~Liu, and L.~Xie, ``Distributed consensus of linear
  multi-agent systems with adaptive dynamic protocols,'' \emph{Automatica},
  vol.~49, no.~7, pp. 1986--1995, 2013.

\bibitem{Dhople14}
S.~V. Dhople, B.~B. Johnson, F.~D\"{o}rfler, and A.~O. Hamadeh,
  ``Synchronization of nonlinear circuits in dynamic electrical networks with
  general topologies,'' \emph{IEEE Transactions on Circuits and Systems---I:
  Regular Papers}, vol.~61, no.~9, pp. 2677--2690, Sep. 2014.

\bibitem{Johnson14}
B.~B. Johnson, S.~V. Dhople, A.~O. Hamadeh, and P.~T. Krein, ``Synchronization
  of nonlinear oscillators in an {LTI} power network,'' \emph{IEEE Transactions
  on Circuits and Systems---I: Regular Papers}, vol.~61, no.~3, pp. 834--844,
  Mar. 2014.

\bibitem{Dorfler13}
F.~D\"{o}rfler and F.~Bullo, ``Kron reduction of graphs with applications to
  electrical networks,'' \emph{IEEE Transactions on Circuits and Systems---I:
  Regular Papers}, vol.~60, no.~1, pp. 150--163, Jan. 2013.

\bibitem{Isidori14}
A.~Isidori, L.~Marconi, and G.~Casadei, ``Robust output synchronization of a
  network of heterogeneous nonlinear agents via nonlinear regulation theory,''
  \emph{IEEE Transactions on Automatic Control}, vol.~59, no.~10, pp.
  2680--2691, Oct. 2014.

\bibitem{Zhang14}
F.~Zhang, H.~L. Trentelman, and J.~M.~A. Scherpen, ``Fully distributed robust
  synchronization of networked {L}ur'e systems with incremental
  nonlinearities,'' \emph{Automatica}, vol.~50, no.~10, pp. 2515--2516, Oct.
  2014.

\bibitem{Arcak02}
M.~Arcak and A.~Teel, ``Input-to-state stability for a class of {L}urie
  systems,'' \emph{Automatica}, vol.~38, no.~11, pp. 1945--1949, Nov. 2002.

\bibitem{Matsumoto85}
T.~Matsumoto, L.~O. Chua, and M.~Komuro, ``The double scroll,'' \emph{IEEE
  Transactions on Circuits and Systems}, vol. {CAS}-32, no.~8, pp. 798--818,
  Aug. 1985.

\bibitem{Kennedy93}
M.~P. Kennedy, ``Three steps to chaos---{P}art {II}: {A} {C}hua's circuit
  primer,'' \emph{IEEE Transactions on Circuits and Systems---I: Fundamental
  Theory and Applications}, vol.~40, no.~10, pp. 657--674, Oct. 1993.

\end{thebibliography}
% -----------------------------------------------------------------------------------------------------------------------------------------

\end{document}